\documentclass[11pt]{article}
\pdfoutput=1
\usepackage{jheppub}
\usepackage{amsmath}
\usepackage{bm}
\usepackage{slashed}
\usepackage{graphicx}
\usepackage{tabularx}
\usepackage{diagbox}
\usepackage{slashed}
\usepackage{youngtab} 
\usepackage{comment}
\usepackage{cancel}
\usepackage[normalem]{ulem}

\newcommand{\tr}{\mathrm{tr}\,}

\newlength{\dummysp}
\newcommand{\beq}{\begin{eqnarray}}
\newcommand{\eeq}{\end{eqnarray}}

\newcommand{\gappeq}{\mathrel{\rlap {\raise.5ex\hbox{$>$}}
{\lower.5ex\hbox{$\sim$}}}}
\newcommand{\lappeq}{\mathrel{\rlap{\raise.5ex\hbox{$<$}}
{\lower.5ex\hbox{$\sim$}}}}

\newcommand{\ben}{\begin{enumerate}}
\newcommand{\een}{\end{enumerate}}

\newcommand{\bit}{\begin{itemize}}
\newcommand{\eit}{\end{itemize}}

\def\[{\left [}
\def\]{\right ]}
\def\({\left (}
\def\){\right )}

\title{New Anomalies, TQFTs, and Confinement\\  in Bosonic Chiral Gauge Theories }
   
\author[a,b]{Mohamed M. Anber,} \author[c,d]{Sungwoo Hong,} \author[e]{Minho Son}
\affiliation[a]{Centre for Particle Theory, Department of Mathematical Sciences, Durham
University, South Road, Durham DH1 3LE, UK}
\affiliation[b]{Department of Physics, Lewis $\&$ Clark College, Portland, OR 97219, USA}
\emailAdd{mohamed.anber@durham.ac.uk}
\affiliation[c]{Department of Physics, The University of Chicago, 5720 S Ellis Ave, Chicago, IL 60637 , USA}
\affiliation[d]{Argonne National Laboratory, 9700 S. Cass Avenue, Lemont, IL 60439, USA}
\emailAdd{sungwooh@uchicago.edu}
\affiliation[e]{Department of Physics, Korea Advanced Institute of Science and Technology,\\
291 Daehak-ro, Yuseong-gu, Daejeon 34141, Republic of Korea\\}
\emailAdd{minho.son@kaist.ac.kr }

\abstract{We study a class of 4-dimensional $SU(N)$ chiral gauge theories with fermions in the 2-index symmetric and antisymmetric representations and classify their infrared phases. The choice $N=4\mathbb{Z}$ corresponds to gauging the fermion number and makes the theory purely bosonic. We examine the most general background  fields of the centers of the gauge, non-abelian flavor, and $U(1)$-axial groups that can be consistently activated, thereby determine the faithful global continuous and discrete symmetries of the theory. This allows us to identify new mixed 0-form/1-form `t Hooft anomalies on both spin and nonspin manifolds. If the theory confines, the absence of composite fermions implies that continuous symmetries must be broken down to anomaly-free subgroups. Anomalies associated with discrete symmetries can be saturated either by breaking the symmetry or by a symmetry-preserving topological quantum field theory (TQFT). The latter, however, is obstructed on spin manifold. The interplay between these features greatly restricts the possible infrared physics. We present two examples that demonstrate our approach. We argue that if the theory confines, the zoo of anomalies and TQFT obstruction greatly restrict the viable infrared condensates. We also discuss the possibility that some theories flow to a conformal fixed point. } 

\begin{document}
\UseRawInputEncoding
\maketitle

\flushbottom

\section{Introduction and Summary}

Classifying the phases of quantum field theory (QFT) is of paramount importance to both high energy and condensed matter physics. Among the landscape of QFT, 4-dimensional asymptotically free chiral gauge theories stand out as one of the most difficult tasks in this endeavor. In addition to their intractably strong coupling at long distances, this in part stems from the fact that putting these theories on a lattice and taking the continuum limit, hoping to learn about their infrared (IR) phases, is still far from being straightforward (see \cite{Poppitz:2010at} for a review).

't Hooft anomaly matching conditions~\cite{tHooft:1979rat} (see also  \cite{Frishman:1980dq,Coleman:1982yg}), or 't Hooft anomalies for short, is among a very few tools available that can shed light on the aforementioned class of theories. It has played a pivotal role in revealing many properties of QFT such as Seiberg dualities \cite{Seiberg:1994pq, Intriligator:1995id} (see \cite{Intriligator:1995au} for a review) and preons \cite{Bars:1982zq} since it was introduced in the 80's.

Given a global symmetry $G$ of QFT, we may try to introduce a background gauge field of $G$. If the symmetry is obstructed, i.e., it becomes anomalous, we say that the theory has an 't Hooft anomaly. Anomalies are renormalization group invariants, and thus, they have to be matched between the ultraviolet (UV) and IR, which imposes constraints on the IR phases of asymptotically free gauge theories. The anomalies of continuous symmetries are matched in the IR either by  breaking the symmetries spontaneously or by composite massless fermions. In both cases the anomalies are matched by a massless spectrum. Anomalies of discrete symmetries, on the other hand, are matched by  breaking the discrete symmetry and forming domain walls, by gapless modes, or by symmetry-preserving TQFT. Another scenario that can match all anomalies is that the theory flows to a conformal field theory (CFT) in the IR. Thus, a trivially gapped vacuum is always excluded in a theory with 't Hooft anomalies.

It was realized in \cite{Gaiotto:2014kfa} that the symmetries and their anomalies can be generalized in a non-trivial way. The authors of \cite{Gaiotto:2014kfa} introduced the concept of generalized global symmetries (GGS). These are p-form global symmetries that act on p-dimensional objects.  Like 0-form symmetries, GGS have Ward identities, organize the spectrum of a theory into representations of the symmetry, and may become anomalous if one tries to gauge them. Therefore, they can be used to impose additional constraints on the IR spectrum of gauge theories.  One of the first applications of the new anomalies was the study of pure $SU(N)$ Yang-Mills theory at $\theta=\pi$ \cite{Gaiotto:2017yup}. This theory enjoys a 1-form $\mathbb Z_N^{(1)}$ center symmetry that acts on Wilson line operators. In addition, at $\theta=\pi$ the theory is time-reversal invariant. It was shown in \cite{Gaiotto:2017yup} that for $N$ even there is a mixed 't Hooft anomaly between the 1-form center and time-reversal symmetries. The matching of this anomaly in the IR means that either the theory breaks the time-reversal symmetry, flows to a Coulomb phase, or that the theory admits a time-reversal preserving TQFT. 

This example illustrates the power of 't Hooft anomalies in imposing tight constraints on the IR phases of strongly-coupled phenomena. Yet, it also shows that the matching conditions allow for a few scenarios rather than pinning down the unique behavior of a given theory. Therefore, it is desirable to have as many anomalies as the theory may possess and as many obstruction conditions in matching those anomalies as possible, hoping that they form sufficiently severe constraints and ultimately lead to a unique possibility for the IR phase.  

The higher-form anomalies, anomalies associated with higher-form symmetries, were taken seriously by many authors, which led to a plethora of papers that examined different aspects of QFT (see~\cite{Tanizaki:2018wtg,Benini:2018reh,Anber:2018iof, Anber:2018jdf, Anber:2018xek, Poppitz:2019fnp,Choi:2018tuh,Komargodski:2017smk,Shimizu:2017asf,Komargodski:2017dmc,Kikuchi:2017pcp,Anber:2019nfu,Aitken:2018kky,Tanizaki:2018xto,Sulejmanpasic:2018upi,Tanizaki:2017mtm,Cox:2021vsa,Poppitz:2020tto} for a non-comprehensive list). In particular, generalizing the 1-form anomalies to also include anomalies of the center of flavors or baryon number was attempted in \cite{Shimizu:2017asf,Tanizaki:2018wtg}. The most general anomalies that involve the color center, flavor center, and baryon number for vector-like theories with fermions in arbitrary representations were studied in \cite{Anber:2019nze,Anber:2020xfk,Anber:2020qzb,Anber:2021lzb}. These anomalies were dubbed the baryon number-flavor-color (BCF) anomalies, and can be envisaged by examining the action of a 0-form symmetry on the fermion content in the presence of 2-form background fields of the centers of the color, flavor, and $U(1)_B$ baryon number. In Ref.~\cite{Anber:2020gig}, the BCF anomalies were further examined in a systematic way on nonspin manifolds. One of the consequences of the BCF anomalies, on spin or nonspin manifolds, is to exclude the possibility that anomalies are saturated in the IR entirely by a set of composite fermions. Either the composites do not form or if they form, then they have to be accompanied by the breaking of discrete chiral symmetries and/or symmetry-preserving TQFTs.  

Another development that has been made is the possibility of anomaly matching by means of symmetry-preserving TQFT \cite{Witten:2016cio, Wang:2017loc, Tachikawa:2017gyf, Wan:2018djl, Cordova:2019bsd, Cordova:2019jqi}. The question is if finite order anomalies, e.g.~$\mathbb{Z}_n$ anomalies, can be saturated by a TQFT when the symmetry is not spontaneously broken. For instance, it was shown in \cite{Cordova:2019bsd} that the mixed anomaly between the 1-form center and 0-form time-reversal symmetries of pure $SU(N)$ Yang-Mills theory at $\theta=\pi$ (for $N$ even) cannot possibly be reproduced by a TQFT since such a TQFT is either not consistent with reflection positivity or does not preserve the symmetry. Likewise, certain exotic phases proposed for 4$d$ adjoint QCD \cite{Bi:2018xvr} were excluded by the absence of the required TQFT \cite{Cordova:2019bsd, Wan:2018djl}.

Chiral gauge theories were examined recently in the light of the traditional (0-form) and higher-form anomalies by various authors \cite{Bolognesi:2017pek,Bolognesi:2019wfq,Bolognesi:2019fej,Bolognesi:2020mpe,Bolognesi:2021yni,Bolognesi:2021hmg,Smith:2021vbf,Sulejmanpasic:2020zfs} (see also  \cite{Csaki:2021xhi,Csaki:2021aqv} for another new direction to study phases of chiral gauge theories). Apparently, these works did not make use of the most general setup introduced in~\cite{Anber:2019nze,Anber:2020gig}, and therefore, new anomalies were missed.

In this paper, we scrutinize a certain class of bosonic chiral gauge theories: these are theories with fermions in the 2-index symmetric and 2-index  anti-symmetric representations of $SU(N)$, with all gauge invariant operators being bosonic. These theories enjoy various continuous and discrete symmetries which make them a prefect playground to examine the interplay between the traditional and newly discovered anomalies. The interplay between the various anomalies as well as the obstruction to symmetry-preserving TQFTs enable us to make nontrivial statements about the IR phases. 
We summarize our findings below.
\begin{itemize}

\item We restrict our discussion to the case $N = 4\mathbb{Z}$. This choice  guarantees that all gauge-invariant operators are bosonic, i.e., the fermion number $(-1)^F$ is gauged. The absence of composite massless fermions plays an important role in limiting the IR options.
These theories admit continuous and discrete chiral symmetries. For the purpose of this summary, and to elucidate subsequent points,  we especially point out to the axial $U(1)_A$ and  discrete chiral $\mathbb Z_p^{d\chi}$ symmetries.  In addition, this class of theories enjoys  a $\mathbb Z_2^{(1)}$ 1-form center symmetry that acts on Wilson lines. 

\item We find new `t Hooft anomalies that have been long missed. While the theory has a $\mathbb Z_2^{(1)}$ 1-form color center, we show that the mere activation of the corresponding background flux by itself does not guarantee a mixed 0-form/1-form anomaly. It is necessary to turn on more refined 1-form symmetry backgrounds, which we achieve by exciting combinations of the color center, flavor center and center of $U(1)_A$ symmetries. We call them CFU backgrounds and the associated anomalies the CFU anomalies. These anomalies can be found on both spin and nonspin manifolds.
\item The mixed $\mathbb Z_p^{d\chi}$[CFU] anomaly places extra constraints on the allowed vacuum operators of the confining phase. This anomaly can be matched either by  breaking  $\mathbb{Z}_p^{d\chi}$ or by a $\mathbb{Z}_p^{d\chi}$-preserving TQFT. The latter option is shown to be excluded on a spin manifold.

\item A given vacuum condensate will generally break $U(1)_A$ down to a discrete subgroup $\mathbb{Z}_q\subset U(1)_A$. If $\mathbb{Z}_q$ is anomalous, the anomaly has to be matched by symmetry-preserving TQFT, which, again is shown to be excluded on a spin manifold.

\item In addition to $U(1)_A$ and $\mathbb Z_p^{d\chi}$, the theory enjoys various continuous symmetries. Since fermions are absent in the IR, the vacuum must break such continuous symmetries to non-anomalous subgroups. This requirement puts further constraints on the possible condensates. 
As a byproduct, we also show that the Wess-Zumino-Witten action~\cite{Wess:1971yu, Witten:1983tw}, which saturates the anomalies of continuous symmetries, will also match the mixed $U(1)_A$ anomaly in the CFU background.

\item Applying the above procedure to the entire space of vacuum operators, we show that only a few choices of the operators (mostly of large scaling dimensions) are admitted as viable candidates.

\item There are only a finite number of bosonic chiral theories of the kind we consider here (see Section \ref{tab:2-index_chiral_theory}). In this work, we consider two examples and present a detailed discussion. In a forthcoming work \cite{follow-up}, we will discuss the rest of cases. The first example discussed in Section~\ref{subsec:SU(8)_k4} is $SU(8)$ theory with a single fermion $\psi$ in the 2-index symmetric representation  and three fermions $\chi$ in the complex conjugate 2-index antisymmetric representation. We show that if the theory confines, a single condensate  that matches $\mathbb{Z}_p^{d\chi}$[CFU] and $\mathbb{Z}_q \subset U(1)_A$ anomalies must have a large scaling dimension. Alternatively, anomalies can be matched by the fermion bilinear, provided that it is accompanied  by another higher dimensional condensate. In the latter case, we show that the IR phase is described by a non-linear sigma model associated with the symmetry breaking $(SU(3) \times U(1))/(\mathbb{Z}_4 \times \mathbb{Z}_3) \to SU(2)$ and domain walls from $\mathbb Z_2^{d\chi}$-breaking.

\item The second example is discussed in Section~\ref{subsec:SU(8)_k2}. It is an $SU(8)$ theory with two $\psi$'s and six $\chi$'s. Anomaly matching conditions leave the vacuum operator $\mathcal{O} \sim \psi^4 \chi^8$ as the lowest dimensional admissible operator. Its large scaling dimension suggests that  the theory may not confine. Interestingly, the $3$-loop perturbative analysis also suggests that the theory flows to an IR fixed point.
\end{itemize}

The rest of the paper is organized as follows. In Sections~\ref{The UV Theory} and \ref{Symmetries} we study the UV theory and determine its discrete and continuous symmetries. Then in Section~\ref{Theory on Spin Manifolds}, we formulate the theory on  spin manifolds and turn on the most general color, flavor, and $U(1)_A$ (CFU) background fluxes. This will enable us to find the faithful global symmetry that acts on fermions. Next, in Section~\ref{`t Hooft anomalies: old and new} we list the conventional anomalies the theory admits and  we use the CFU background to detect new 't Hooft anomalies. The next logical step, in Section~\ref{Theory on Nonspin Manifolds}, is to put the theory on a nonspin manifold and, again, turn on the CFU fluxes. This has to be done in a way that renders the fermions well-defined on such manifolds. We, then, show that new anomalies can be detected on nonspin manifolds. In Section~\ref{Matching the Anomalies in the IR}, we discuss the IR matching of the UV anomalies. In particular, we show that theories with $N=4\ell$, for some integer $\ell$, are bosonic, and thus, gauge-invariant fermion operators are not available to match the anomalies.  We also comment on theorems regarding symmetry-preserving TQFTs and describe a universal TQFT that matches anomalies of broken discrete symmetries. Finally, in Section~\ref{subsec:SU(8)_k4} and ~\ref{subsec:SU(8)_k2} we discuss the candidate phases of the bosonic theories and present the two examples mentioned above. A few points that intrude the discussion in the bulk of the paper are presented in Appendices~\ref{The discrete symmetry} to \ref{2loop beta function}.

Throughout the paper we use lower-case letters to denote the dynamical (color) field, e.g., $a_1^c$, $f_2^c$ to denote the 1-form gauge field and the 2-form field strength. While we use upper-case letters to denote nondynamical (background) field, e.g., $F_2^f$ for the 2-form flavor background field.

\section{Chiral Gauge Theories with $2$-index Fermions}
\label{chiral gauge theories with 2-index}

In this section we describe a class of chiral gauge theories whose fermion content consists of 2-index symmetric and 2-index anti-symmetric
representations of the gauge group $SU(N)$. We discuss the global symmetries of the theory in detail and carefully identify the most general possible background fields that can be turned on. The latter will be used to identify the faithful global symmetry and will lead to an additional set of `t Hooft anomalies. This provides a refined probe of the theory and results in extra constraints on the allowed IR phases.

\subsection{The UV Theory}
\label{The UV Theory}

We consider $SU(N)$ gauge theory with $n_\psi$ flavors of left-handed Weyl fermions $\psi$ in the two-index symmetric ($\tiny \yng(2)$) representation and $n_\chi$ flavors of left-handed Weyl fermions $\chi$ in the complex conjugate two-index anti-symmetric ($\tiny \overline{\yng(1,1)}$) representation. The classical Lagrangian is 
\beq
\mathcal{L} = - \frac{1}{4g^2} f^{c\, A}_{\mu\nu} f^{c\,A \mu\nu} - i \bar{\psi} \bar{\sigma}^\mu D_\mu \psi - i \bar{\chi} \bar{\sigma}^\mu D_\mu \chi + \frac{\theta}{32 \pi^2} \epsilon^{\mu\nu\rho\lambda} f^{c\, A}_{\mu\nu} f^{c\,A}_{\rho\lambda}
\eeq
where $A$ is the index of $SU(N)$ and we have suppressed the gauge, spinor and flavor indices for fermions. For future reference, $\psi$ carries two upper color indices and $\chi$ carries $N-2$ upper indices. The latter is equivalent to having $2$ lower indices.  Recalling that the anomaly coefficients of the 2-index fermions are given by $\mathcal{A}_{\tiny \yng(2)} = N+4$ and $\mathcal{A}_{\overline{\tiny \yng(1,1)}} =-( N-4)$, we find that the number of fermions species is constrained to be
\beq
n_\psi = \frac{N-4}{k}, \;\;\; n_\chi = \frac{N+4}{k}
\eeq
where $k$ is a common divisor of $(N-4, N+4)$ and we take $N\geq 5$. This class of theories is asymptotically free provided $11N -\frac{2(N^2-8)}{k}>0$ (note that it does not admit large-$N$ limit). These conditions single out the following set of asymptotically free $2$-index chiral theories:
\begin{equation}
\begin{tabular}{|c|c|c|c|c|c|c|c|c|c|c|}
\hline
$N$ &  $5$ &   $6$ &   $8$ &   $10$ &   $12$  &  $16$  &  $20$ &   $28$ &   $36$ &   $44$ \\
\hline
 $k$ & $1$ &  $1,2$ &  $2,4$ &  $2$ &  $4,8$ &  $4$ &  $4,8$ &  $8$ &  $8$ & $8$\\
\hline
\end{tabular}\,.
\label{tab:2-index_chiral_theory}
\end{equation}

The $\theta$-term appears in the path integral in the form
\beq
{\cal Z} = \sum_n e^{i n \theta}{\cal Z}_n
\label{eq:theta_vacuum}
\eeq
where ${\cal Z}_n$ is the partition function with a fixed instanton number $n$
\beq
n = \frac{1}{8\pi^2} \int_{\mathbb{M}^4} {\rm tr}_\Box \left( f^c_2\wedge f^c_2 \right) = \frac{1}{32\pi^2} \int d^4 x \; \epsilon^{\mu\nu\rho\lambda} f^{c\,A}_{\mu\nu} f^{c\,A}_{\rho\lambda}\,.
\label{eq:instanton_number_SU(N)}
\eeq 
where the trace is taken over the defining representation of the color group with the normalization $\tr_{\Box}\left[t^at^b\right]=\delta^{ab}$ where $t^{a}$ refers to the Lie-algebra generators. The particular form of eq.~(\ref{eq:theta_vacuum}) is required by locality (cluster decomposition) and unitarity. While in the presence of massless fermion we may use a chiral rotation to remove the $\theta$-angle term, it is nonetheless useful to keep it for the discussion of symmetries.

\subsection{Symmetries}
\label{Symmetries}

In this subsection, we discuss the global symmetries of the theory. We primarily focus on identifying the global symmetries without 
paying attention to possible refinements. The correct faithful global symmetry that acts on fermions will be discussed in the next section.

We start with the continuous flavor symmetries. The theory admits the flavor symmetry of $U (n_\psi) \times U(n_\chi)$ ($\simeq SU (n_\psi) \times SU(n_\chi) \times U(1)_\psi \times U(1)_\chi$) that acts on $\psi$ and $\chi$, respectively. 
$\psi$ ($\chi$) transforms in the defining (anti-defining) representation of $SU (n_\psi)$ ($SU(n_\chi)$). 
Of two classical $0$-form $U(1)$ symmetries, $U(1)_\psi \times U(1)_\chi$, only one linear combination survives as a good symmetry at the quantum level. We shall call this anomaly-free $U(1)$ symmetry the \emph{axial} $U(1)_A$. The other combination is broken down to a discrete subgroup by the color instantons (i.e. the ABJ anomaly). Assuming that $\psi$ ($\chi$) carries the $U(1)_A$ charge $q_\psi$ ($q_\chi)$, the $U(1)_A$ rotation by an angle $\alpha$ shifts the $\theta$-angle by
\beq
\theta \to \theta + \left( N_\psi q_\psi + N_\chi q_\chi \right) \alpha,
\eeq
where $N_\psi = n_\psi T_\psi$ and $N_\chi = n_\chi T_\chi$ with Dynkin indices, $T_\psi = T_{\tiny \yng(2)} = N+2$ and $T_\chi = T_{\overline{\tiny \yng(1,1)}} = N-2$.
The anomaly-free condition is achieved for the minimal coprime $U(1)_A$-charges
\beq
q_\psi = - \frac{N_\chi}{r}~, \quad q_\chi = \frac{N_\psi}{r}~,\quad {\rm where}~ \quad r = {\rm gcd} (N_\psi, N_\chi)~, 
\label{eq:U(1)_AF_charges}
\eeq
or any (common) integral multiples of them.

We next move onto the discrete symmetry of the theory. As mentioned above, any linear combinations of $U(1)_\psi \times U(1)_\chi$ other than the direction defined by eq.~(\ref{eq:U(1)_AF_charges}) are explicitly broken by the ABJ anomaly.  Let us consider $U(1)$ with $(\psi, \chi)$-charges equal to $(1,1)$. Under this $U(1)$ transformations, $\theta$ is shifted by
\beq
\theta \to \theta + \left( N_\psi + N_\chi \right) \alpha~,
\eeq
implying that $U(1)$ is broken down to $\mathbb{Z}_{\scriptscriptstyle N_\psi + N_\chi}$. In Appendix~\ref{The discrete symmetry} we show that some of $\mathbb{Z}_{\scriptscriptstyle N_\psi + N_\chi}$ generators are redundant since they can be absorbed in a $U(1)_A$ transformation. The independent symmetry is shown to be 
\beq
\frac{U(1)_A \times \mathbb{Z}_{\scriptscriptstyle N_\psi + N_\chi}}{\mathbb{Z}_{\scriptscriptstyle ( N_\psi + N_\chi ) / r}}~.
\label{eq:U(1)_A_Z_r}
\eeq
Hence, there are only $r$ independent discrete generators\footnote{Importantly, this is insensitive to the $U(1)_A$ charge assignment. This means that multiplying the $U(1)_A$ charges of $\psi$ and $\chi$ by a common integer cannot alter the number of independent generators.}.
The independent discrete rotations can conveniently be represented as a $\mathbb Z_r$ transformation that acts solely\footnote{ Alternatively, we could choose $\mathbb Z_r$ to act on $\psi$.} on $\chi$:
\begin{eqnarray}
(\psi, \chi)\rightarrow (\psi,e^{i \frac{2\pi \ell}{r}} \chi)\,\quad \ell=0,1,2,...,r-1\,,
\label{eq:Z_r_transformation}
\end{eqnarray}
and we refer the reader to Appendix \ref{The discrete symmetry} for a derivation.  However, we must stress that the ultimate genuine discrete chiral symmetry can be a proper subgroup $\mathbb{Z}_p^{d\chi} \subset \mathbb{Z}_r$ and is pinned down by taking into account any further redundancies. In particular, a combination of the generators of the abelian subgroups of the theory, i.e.~a combination of the 0-form center of the color group (see below for a further discussion), center of flavor $SU(n_f)$, and $\mathbb{Z}_2^{\scriptscriptstyle \rm L}$ Lorentz, can reproduce a subset of the generators of $\mathbb{Z}_r$. We check this possibility for the examples we study in Sections~\ref{subsec:SU(8)_k4} and ~\ref{subsec:SU(8)_k2}. The correct identification of the genuine discrete symmetry is crucial because only then one obtains the correct number of degenerate vacua (assuming the vacuum breaks the discrete symmetry).

In addition, the theory has several 0-form discrete symmetries. First of all, there is a 0-form color center\footnote{In fact, this is a gauge redundancy rather than a global symmetry.} $\mathbb{Z}_{N/{\rm gcd} (N,n)}$ ($n=2$ is the $N$-ality of fermions) that acts on $\psi$ and $\chi$ as a phase rotation. Likewise, there are 0-form centers of  the flavor groups, $Z \left( SU(n_\psi) \right) = \mathbb{Z}_{n_\psi}$ and $Z \left( SU(n_\chi) \right) = \mathbb{Z}_{n_\chi}$, that act on $\psi$ and $\chi$, respectively. There is also a $\mathbb{Z}_2^{\scriptscriptstyle \rm L}\equiv (-1)^F$ discrete subgroup of the Lorentz group, which is simply a $2\pi$-rotation, and thus, it is contained in the connected part of the group. It, nonetheless, can play an important role when we discuss background fields on nonspin manifolds. 

Finally, the theory has a  $\mathbb Z^{(1)}_{\scriptscriptstyle {\rm gcd} (N, n)}$ $1$-form symmetry that acts on the Wilson line operators. For $N$ even we have $\mathbb{Z}_2^{(1)}$ 1-form symmetry which acts on the Wilson operator in the fundamental representation, while for $N$ odd we do not have any 1-form symmetry.

We summarize the above discussion by giving the (naive) global symmetry (not including spacetime symmetry):
\begin{eqnarray}
G= SU (n_\psi) \times SU (n_\chi) \times U (1)_A \times \mathbb{Z}_r \times \mathbb{Z}^{(1)}_{\scriptscriptstyle {\rm gcd} (N, 2)} \label{eq:naive global symmetry}\,,
\end{eqnarray}
where $SU (n_\psi) \times SU (n_\chi) \times U (1)_A \times \mathbb{Z}_r$ act on the fermions, while $ \mathbb{Z}^{(1)}_{\scriptscriptstyle {\rm gcd} (N, 2)}$ acts on the fundamental Wilson lines.  We emphasize that this expression is still rather imprecise (as there may be additional redundancies) and it will be further refined in the next section.

\subsection{Theory on Spin Manifolds}
\label{Theory on Spin Manifolds}

Now, we discuss the most general background fluxes of $SU(N)\times SU (n_\psi) \times SU (n_\chi) \times U(1)_A$ that can be activated on a spin manifold. This procedure enables us to determine the faithful group that acts on the matter fields and, at the same time, gives rise to the most stringent set of 't Hooft anomalies that must be matched between the UV and IR.

\subsubsection{Background gauge fields: color-flavor-$U(1)$ (CFU) fluxes}
\label{Background Gauge Fields and Global Structure of the Symmetry Group}

Consider a group $G$ with a non-trivial center $Z (G)$. We want to turn on a background gauge field of $G/ Z(G)$.  To this end, consider a gauge theory on a closed four-dimensional manifold $\mathbb{M}^4$, and let $\phi$ be a matter field transforming under $G$  in a representation $\cal R$. Then, define a $G$-bundle on $\mathbb{M}^4$. Given coordinate charts $\{ U_i \}$ of the base manifold $\mathbb{M}^4$ and sections $\phi_i$ on $U_i$, the matter fields on the double overlap $U_{ij} = U_i \cap U_j$ are related by
\beq
\phi_i = {\cal R} (g_{ij} ) \phi_j, \;\; g_{ij} : U_{ij} \to G~,
\eeq
where the transition function ${\cal R} (g)$ denotes the group element $g$ in $\cal R$. On the triple overlap, $U_{ijk} = U_i \cap U_j \cap U_k$, the cocycle condition (which is a compatibility condition) should be satisfied:
\beq
{\cal R}(g_{ij}) \circ {\cal R}(g_{jk}) \circ {\cal R}(g_{ki}) = {\cal R}(\mathbf{1}_G)~,
\label{general cocycle}
\eeq
where $\mathbf{1}_G$ is the identity operator in $G$. For the gauge field $A$, the compatibility condition on a double overlap is given by
\beq
A_i = g_{ij} \left( A_j - i d \right) g_{ij}^{-1}~,
\eeq
where $A_i$ denotes the gauge field on the chart $U_i$ and $g_{ij}$ is a $G$-valued function on $U_{ij}$, in the defining representation of $G$, with the property $g_{ji} = g_{ij}^{-1}$. The cocycle condition on the triple overlaps is
\beq
g_{ij} \circ g_{jk} \circ g_{ki} = \mathbf{1}_G~.
\label{ordinary cocycle}
\eeq
This cocycle condition is stronger\footnote{According to our discussion, the condition (\ref{ordinary cocycle}) can be relaxed to $g_{ij} \circ g_{jk} \circ g_{ki} =e^{i2\pi \frac{n_{ijk}}{N}}$, for integers $n_{ijk}$, in the case of pure $su(N)$ Yang-Mills theory or Yang-Mills theory with adjoint fermions. This cocycle condition describes the $SU(N)/\mathbb Z_N$ bundle. This fact was first realized by 't Hooft \cite{tHooft:1979rtg}. } than (\ref{general cocycle}). As we will see momentarily, (\ref{ordinary cocycle}) can be relaxed when we consider  $G/Z(G)$ instead of  $G$ bundle. 

Now, we are in a position to construct the principal bundle of $G/Z(G)$. As an intermediate step, however,  we introduce $\hat{G} \equiv \frac{G \times U(1)}{Z(G)}$ as an auxiliary bundle. The principal bundle associated with $\hat{G}$ satisfies a modified compatibility condition. The identification by the center $Z(G)$ means the following. Let $(g,u) \in G \times U(1)$. Then, elements of $\hat{G}$ are defined up to an equivalence relation
\beq
(g, u) \sim (g z, z^{-1} u), \;\; z \in Z(G)~.
\eeq
Therefore, the $\hat{G}$-bundle is defined by
\beq
(g, u)_{ij} \circ (g, u)_{jk} \circ (g, u)_{ki} = (z, z^{-1}) \sim (1,1)~.
\label{eq:cocycle_condition_with_center}
\eeq
We observe that the $G$- and $U(1)$-bundles are not well-defined independently (in fact, $g_{ij} \circ g_{ji} \circ g_{ki} = z \in Z(G)$ shows that the center $Z(G)$  is gauged \cite{Kapustin:2014gua}). Only the combination defines a  sensible structure, the $\hat{G}$-bundle.

We can go one step further and finally construct the $G/ Z(G)$-bundle by projecting the $U(1)$ onto the $Z(G)$ direction, which often can be achieved by imposing a 1-form gauge symmetry. Let us consider $G=SU(N)$ (with $Z(SU(N)) = \mathbb{Z}_N$) to demonstrate this idea\footnote{Again, this is the example of pure Yang-Mills theory or Yang-Mills theory with adjoint fermions.}. We first promote $G \to \hat{G} = \frac{SU(N) \times U(1)}{\mathbb{Z}_N} = U(N)$. Denoting the $U(N)$ 1-form gauge field as $\hat{a}_1$ and its 2-form field strength as $\hat{f}_2$, we want to study
\beq
S =  \frac{1}{g^2} \int {\rm tr} \left( \hat{f}_2 \wedge \star \hat{f}_2 \right) +  \frac{i}{2\pi} \int F_2 \wedge {\rm tr} \left( \hat{f}_2 \right) +  \frac{i \theta}{8\pi^2}  \int {\rm tr} \left( \hat{f}_2 \wedge \hat{f}_2 \right)~.
\eeq 
We have introduced a 2-form Lagrange multiplier  $F_2$ in order to project out the trace part of the $U(N)$ field. While this successfully removes one local degree of freedom (recall $U(N)$ has one more local degree of freedom than $SU(N)$ or $SU(N) / \mathbb{Z}_N$), it does not yield the desired theory, $SU(N) / \mathbb{Z}_N$. The issue is that while the `t Hooft flux of $SU(N) / \mathbb{Z}_N$  takes values in $H^2 (X, \mathbb{Z}_N)$, that of $U(N)$ theory is $H^2 (X, \mathbb{Z})$-valued. This problem can be fixed by postulating a 1-form gauge symmetry that acts on $\hat a_1$ as:
\beq
\hat{a}_1 \to \hat{a}_1 - \lambda_1 \mathbf{1}~,
\eeq
where the 1-form gauge symmetry parameter $\lambda_1$ itself is a $U(1)$ gauge field. 
Writing $\hat{a}_1 = a_1 + \frac{1}{N} \hat{A}_1 \mathbf{1}$ with $a_1$ and $\hat{A}_1$ being the $SU(N)$ and $U(1)$ gauge fields, respectively, the 1-form gauge transformation may be restated as
\beq
a_1 \to a_1, \;\;\;\; \hat{A}_1 \to \hat{A}_1 - N \lambda_1~.
\eeq
The $U(N)$ field strength is given by $\hat{f}_2 = f_2 + \frac{1}{N} d \hat{A}_1 \mathbf{1}$ and transforms as $\hat{f}_2 \to \hat{f}_2 - d \lambda_1 \mathbf{1}$. This shows that the above action is not invariant. To remedy the problem, we introduce a 2-form gauge field $B_2$ which shifts under the 1-form transformation as $B_2 \to B_2 - d \lambda_1$. It is then clear that the combination $\hat{f}_2 - B_2 \mathbf{1}$ is invariant under the 1-form gauge transformation. Hence, the invariant action is given by
\beq
\begin{split}
S = & \frac{1}{g^2} \int {\rm tr} \left[ \left( \hat{f}_2 - B_2 \mathbf{1} \right) \wedge \left( \hat{f}_2 - B_2 \mathbf{1} \right) \right] + \frac{i}{2\pi} \int F_2 \wedge {\rm tr} \left( \hat{f}_2 - B_2 \mathbf{1} \right) 
\\
& +  \frac{i \theta}{8\pi^2}  \int {\rm tr} \left[ \left( \hat{f}_2 - B_2 \mathbf{1} \right) \wedge \left( \hat{f}_2 - B_2 \mathbf{1} \right) \right]~.
\end{split}
\eeq
Let us now show that the instanton number of this theory is indeed $\mathbb{Z}_N$-valued. To see this, one notes that ${\rm tr} ( \hat{f}_2 ) = d \hat{A}_1 \equiv \hat{F}_2$. Using this expression, the Lagrange multiplier term becomes
\beq
\frac{i}{2\pi} \int F_2 \wedge \left( d \hat{A}_1 - N B_2 \right)~.
\label{eq:BF_action}
\eeq
One recognizes that this is just the BF theory describing a $\mathbb{Z}_N$ gauge theory. It imposes the constraint $B_2 = \frac{1}{N} d \hat{A}_1$. Then, the $\theta$-term can be arranged into a form
\beq \label{eq:fractional_instanton_SU(n)_center}
\begin{split}
S_\theta &= \frac{i \theta}{8\pi^2}  \int  {\rm tr} \left( \hat{f}_2 \wedge \hat{f}_2 \right) - N B_2 \wedge B_2
\\[3pt]
&= \frac{i \theta}{8\pi^2}  \int {\rm tr} \left( \hat{f}_2 \wedge \hat{f}_2 \right) - {\rm tr} \left( \hat{f}_2 \right) \wedge {\rm tr} \left( \hat{f}_2 \right) + \frac{i \theta}{8\pi^2}  \int \frac{N-1}{N} \hat{F}_2 \wedge \hat{F}_2 
\\[3pt]
&= i \theta n_{\scriptscriptstyle SU(N)} + i \theta \left( \frac{N-1}{2N} \right) \int w_2 \wedge w_2~,
\end{split}
\eeq 
where $\hat{F}_2 = d \hat{A}_1$ and $n_{\scriptscriptstyle SU(N)}$ is the $SU(N)$ instanton number introduced in eq.~(\ref{eq:instanton_number_SU(N)}), which is integer valued. In the last line of eq.~(\ref{eq:instanton_number_SU(N)}), we defined $w_2 = \frac{\hat{F}_2}{2\pi} = N \frac{B_2}{2\pi}$. One notes that $\int w_2 \wedge w_2  \in 2 \mathbb{Z}$ on a spin-manifold. The last form of the presentation makes it especially clear that the second term describes a fractional `t Hooft flux valued in $H^2 (X, \mathbb{Z}_N)$ and the resulting theory is that of $SU(N) / \mathbb{Z}_N$. In what follows, we will use this result directly when we consider our chiral theory and turn on background fluxes of the centers of the color and flavor groups.

We also discuss turning on the center background field of a $U(1)$ theory since it will play an important role in chiral gauge theories.
A $U(1)$ theory has electric 1-form and magnetic 1-form symmetries (both are $U(1)$ 1-form symmetries with a mixed anomaly). 
Here, we turn on the background field for the electric 1-form center (see Footnote \ref{explaining U(1) background} for an example and further explanation of this point.). An invariant topological term under a 1-form gauge transformation $A_1 \to A_1+ \alpha \lambda_1$ ($\alpha \in \mathbb{R}$ and $\oint_{\gamma_1} \frac{\lambda_1}{2\pi} \in \mathbb{Z}$) can be written as
\beq
 \frac{i \theta}{8\pi^2} \int  \left( F_2 - B_2 \right) \wedge  \left( F_2 - B_2 \right)~.
\label{eq:U(1)_center_flux}
\eeq
Invariance under the 1-form gauge transformation requires that  the 2-form center background gauge field transforms as $B_2 \to B_2 + \alpha d \lambda_1$. While the $U(1)$ instanton term vanishes on $S^4$ (since $\pi_3 (U(1)) = 0$, there are no $U(1)$ instantons), it can nonetheless be non-zero when the manifold supports non-trivial 2-cycles.

Using all the ingredients discussed above, we can discuss the most general background fluxes that can be activated in our chiral theory. We take $\mathbb M^4$ to be a spin manifold, or it admits a spin structure such that the spinors $\psi$ and $\chi$ are well-defined on $\mathbb M^4$.
We also define a principal bundle of the continuous part of the global symmetry $G$ (see eq.~(\ref{eq:naive global symmetry})) on $\mathbb M^4$ and take the transition functions of $G$ to act on fibers by left multiplication. Spinors are sections of the bundle and we use the notations $\psi_i$ and $\chi_i$ for their values on a local patch $U_i$.  We denote the transition functions of the color $SU(N)$, (non-abelian) flavor, and $U(1)_A$ group as $g$, $f$, and $u$, respectively, along with the proper superscript to distinguish those of $\psi$ and $\chi$. 

On the double overlap $U_{ij} = U_i \cap U_j$ we have
\beq
\psi_i = (g^\psi, f^\psi, u^\psi)_{ij} \, \psi_j~,\quad \chi_i = (g^\chi, f^\chi, u^\chi)_{ij} \, \chi_j~.
\eeq
The exact form of the cocycle condition on the triple overlap depends on the allowed background gauge fields (similar to the $\hat{G}$-bundle in eq.~(\ref{eq:cocycle_condition_with_center})). We turn on background gauge fields for centers of the gauge, flavor, and $U(1)_A$ groups and determine the most general combination compatible with the cocycle condition. With non-trivial center fluxes, the resulting form of the cocycle conditions is
\beq
\left( g^\psi, f^\psi, u^\psi \right)_{ij} \circ \left( g^\psi, f^\psi, u^\psi \right)_{jk} \circ \left( g^\psi, f^\psi, u^\psi \right)_{ki} = \left( z_c, z_f, z_u \right) \quad {\rm with} \quad  z_c z_f z_u = 1~,
\eeq
where $z$'s refer to the center elements: $z_c \in \mathbb{Z}_{N/{\rm gcd}(N,2)}$, $z_f \in \mathbb{Z}_{n_\psi}$, and $z_u \in U(1)_A$.
The condition $z_c z_f z_u = 1$ is required for the equivalence relation,\footnote{For given group elements $(g,f,u) \in SU(N) \times SU(N_f) \times U(1)$, this is equivalent to 
\begin{equation}
(g, f, u) \sim (z_c g, z_f f, z_u u).
\label{eq:equivalence relation 2}
\end{equation}
}
\beq
\left( z_c, z_f, z_u \right) \sim (1,1,1)~,
\label{eq:equivalence relation}
\eeq
which is needed to obtain the correct compatibility condition. Similar expressions hold for the cocycle condition of $\chi$. 
The possible background fields are therefore determined by solving the following consistency equations:
\begin{eqnarray}
&& \psi \; : \;\; \underbrace{e^{i2\pi\frac{2m}{N}}}_{z_c} \underbrace{e^{i2\pi \frac{pk}{N-4}}}_{z_f} \underbrace{e^{-i 2\pi s \frac{(N+4)(N-2)}{kr}}}_{z_u} = 1 \label{eq:detailed cocycles psi}~, 
\\[3pt]
&& \chi : \;\;  \underbrace{e^{-i2\pi\frac{2m}{N}}}_{z_c} \underbrace{e^{-i2\pi \frac{p'k}{N+4}}}_{z_f} \underbrace{e^{i 2\pi s\frac{(N-4)(N+2)}{kr}}}_{z_u}=1~,
\label{eq:detailed cocycles chi}
\end{eqnarray}
where $m \in \mathbb \mathbb{Z}_{N/{\rm gcd}(N,2)}$, $p\in \mathbb Z_{\frac{N-4}{k}}$, $p'\in \mathbb Z_{\frac{N+4}{k}}$ and $s$ is a $U(1)_A$ parameter.
The factor of 2 that appears in $z_c$ is the $N$-ality of $\psi$ and $\chi$, and we have written the number of flavors $n_\psi$ and $n_\chi$ explicitly in the $z_f$ factors. 
We shall call the background fluxes associated with the non-trivial solutions of eqs.~(\ref{eq:detailed cocycles psi}) and (\ref{eq:detailed cocycles chi}) the Color-Flavor-$U(1)$  `t Hooft fluxes, or  CFU  fluxes for short.

There exist multiple solutions to the CFU consistency equations. As we show below, the class of solutions forms a discrete group (or a direct product of discrete groups), which can be used to single out the faithful global symmetry that acts on fermions. We also note that while the sole activation of the $\mathbb Z_2$ color-center is always possible for $N$ even as a special solution, this background may not lead to new or the most refined anomalies. On the contrary, as will be discussed in Section~\ref{subsec:SU(8)_k4} and ~\ref{subsec:SU(8)_k2}, the discrete chiral symmetry $\mathbb{Z}_p^{d\chi} \subset \mathbb{Z}_r$ exhibits a non-trivial anomaly in more general CFU-backgrounds, an effect that  is absent if we turn on only the color-center background. 

Once a non-trivial solution is found, we can calculate the topological charges associated with center fluxes. For this we note that the effect of a center flux is to modify the instanton number. Using $Q$ with proper subscript to denote the fractional instanton number (or topological charges), from eqs.~(\ref{eq:fractional_instanton_SU(n)_center}) and (\ref{eq:U(1)_center_flux}), we have
\beq
&& Q_c = \left( 1 - \frac{1}{N} \right) \int \frac{w_2 (c) \wedge w_2 (c)}{2} = m_1 m_2 \left( 1 - \frac{1}{N} \right)~, \label{eq:Q_c}
\\[3pt]
&& Q_\psi = \left( 1 - \frac{1}{n_\psi} \right) \int \frac{w_2 (\psi) \wedge w_2 (\psi)}{2} = p_1 p_2 \left( 1 - \frac{k}{N-4} \right)~, \label{eq:Q_psi}
\\[3pt]
&& Q_\chi = \left( 1 - \frac{1}{n_\chi} \right) \int \frac{w_2 (\chi) \wedge w_2 (\chi)}{2} = p^\prime_1 p^\prime_2 \left( 1 - \frac{k}{N+4} \right)~. \label{eq:Q_chi} 
\\[3pt]
&& Q_u = \frac{1}{8\pi^2} \int  \left( F_2 - B_2 \right) \wedge  \left( F_2 - B_2 \right) = (n_1 - s_1) (n_2 - s_2), \;\; n_1, n_2 \in \mathbb{Z}~. \label{eq:Q_u}
\eeq
We used the notation $w_2 (i), i=c, \psi, \chi$ to represent the center background 2-form fields for the color group, $\psi$-flavor-center group and $\chi$-flavor-center group, respectively~\footnote{The same results can also be obtained by explicit construction of center background gauge fields.
As a concrete example, we take $\mathbb M^4$ to be a symmetric four-torus $\mathbb T^4$, with a cycle length $L$, and turn on background CFU-fluxes ('t Hooft fluxes) along two independent cycles. For example, we can turn on fields in the  $x^1x^2$ and $x^3x^4$ planes of $\mathbb T^4$ as follows (see \cite{Anber:2019nze} for details):
\begin{eqnarray}\label{eq:t Hooft fluxes_explicit}
\begin{split}
A_1^c&=\left(\frac{2\pi m_1}{L^2}\bm H^c \cdot \bm \nu^c \right)x^2\,, \quad A_2^c=0\,,\quad
A_1^\psi =\left(\frac{2\pi p_1}{L^2}\bm H^\psi \cdot \bm \nu^\psi \right)x^2\,, \quad A_2^\psi =0~,
\\[3pt]
A_1^\chi &=\left(\frac{2\pi p'_1}{L^2}\bm H^\chi \cdot \bm \nu^\chi \right)x^2\,, \quad A_2^\chi =0\,,\quad
A_1^u =\frac{2\pi s_1}{L^2} x^2\,, \quad A_2^u =0~,
\end{split}
\end{eqnarray}
where $\bm H$ are the Cartan generators (for example, $\bm H^c$ for color group is a vector matrix with $N-1$ components) and $\bm \nu$ are the weights, both are taken in the defining representations of the respective groups.  Similar expressions hold for the fields in the $x^3x^4$ planes after replacing $1\rightarrow 3$ and $2\rightarrow 4$, including the integers $m,p,p'$ and the $U(1)$ parameter $ s$. Notice that $m,p,p', s$ in the $x^1x^2$ and $x^3x^4$ planes need not be the same. Yet, both sets must solve the consistency equations (\ref{eq:detailed cocycles psi}) and (\ref{eq:detailed cocycles chi}). Using the fluxes (\ref{eq:t Hooft fluxes_explicit}), one can readily calculate the topological charges $Q=\frac{1}{8\pi^2} \int_{\mathbb T^4} \mbox{tr}_{\Box}\left[F \wedge F\right]$, which in fact agree with eqs.~(\ref{eq:Q_c}) -- (\ref{eq:Q_u}), as expected.
}. $B_2$ is the $U(1)_A$ center background fields. Let us explain the meaning of the integers $m_1,m_2,p_1,p_2,p'_1,p'_2$.
Consider the 4$d$ spin manifold $\mathbb M^4 = \mathbb M^2 \times {\mathbb M^\prime}^2$ with $\mathbb M^2$ and $\mathbb M'^2$ are either $\mathbb S^2$ or $\mathbb T^2$.  For $SU(n)$ centers ($SU(n)$ account for either color or flavor groups), we consider a background configuration which is a sum of $a$ units of fluxes piercing $\mathbb M^2$ with zero flux through $\mathbb M'^2$ and zero flux through $\mathbb M^2$ with $b$ units of fluxes piercing $\mathbb M'^2$, i.e.~$w_2 = a (\mathbb M^2) \times 0 ({\mathbb M'}^2) + 0 (\mathbb M^2) \times b ({\mathbb M'}^2)$. This leads to
\beq
\int_{\mathbb M^2 \times {\mathbb M'}^2} w_2 \wedge w_2 = 2 ab\,,
\eeq
and hence eqs.~(\ref{eq:Q_c}) -- (\ref{eq:Q_chi}).
For $U(1)_A$ we repeat the same exercise: we take the background field to be the sum of two terms, each term having a non-zero flux piercing either  $\mathbb M^2$ or $\mathbb M'^2$. Then, $n_1, n_2$ are integer-valued fluxes of $F_2$ on $\mathbb M^2$ and $\mathbb M'^2$, respectively, i.e.~$\int \frac{F_2}{2\pi} = n_{1,2}$.
Similarly, $s_1, s_2$ are the fractional $U(1)_A$ fluxes of $B_2$ along $\mathbb M^2$ and $\mathbb M'^2$.

As a consistency check on our construction, we can also calculate the Dirac-index of both $\psi$ and $\chi$:
\beq
&& {\cal I}_{\psi} = n_\psi T_{\psi} Q_c + \mbox{dim}_\psi  Q_\psi + \mbox{dim}_\psi n_\psi q_{\psi}^2 Q_u~, \label{eq:Dirac index psi}
\\[3pt]
&& {\cal I}_{\chi} = n_\chi T_{\chi} Q_c + \mbox{dim}_{\chi}  Q_\chi + \mbox{dim}_{\chi}  n_\chi q_\chi^2 Q_u~,
\label{eq:Dirac index chi}
\eeq
where $\mbox{dim}_\psi=\frac{N(N+1)}{2}$ and $\mbox{dim}_{\chi}=\frac{N(N-1)}{2}$ are the dimensions of the fermion representations under $SU(N)$. The Dirac-index counts the number of the Weyl zero modes in the background of the topological charges eqs.~(\ref{eq:Q_c}) -- (\ref{eq:Q_u}). Hence, the consistency of our construction demands the integrality of the index. This can be explicitly checked for any solutions to eqs.~(\ref{eq:detailed cocycles psi}) and (\ref{eq:detailed cocycles chi}).

Finally, we answer the question that pertains to the faithful group that acts on the fermions in our theory. Finding the faithful group amounts to identifying the elements that are  common to the color, flavor, and $U(1)_A$ groups. This identification is exactly the equivalence relation (\ref{eq:equivalence relation}), which in turn leads to the consistency equations (\ref{eq:detailed cocycles psi}) and (\ref{eq:detailed cocycles chi}).  First, we discuss the identification of an element of the center of the color group and an element of $U(1)_A$. If we can find a solution of the consistency equations   (\ref{eq:detailed cocycles psi}) and (\ref{eq:detailed cocycles chi}) for $m \neq 0$ (mod $N$) and $s \neq 0$, setting $p=p'=0$, then the faithful 0-form global symmetry acting on fermions should be modded by $\mathbb Z_{N/\scriptsize\mbox{gcd}(N,2)}$. We can obtain an algorithmic generalization of this finding as follows. We search for the complete set of solutions to eqs. (\ref{eq:detailed cocycles psi}) and (\ref{eq:detailed cocycles chi}) and pin down the discrete group (or the direct product of discrete groups) that generate the full set of solution. Modding the continuous part of (\ref{eq:naive global symmetry}) by this discrete group gives the faithful global symmetry that acts on fermions. For example, if solutions to   eqs. (\ref{eq:detailed cocycles psi}) and (\ref{eq:detailed cocycles chi}) exits for all independent integers $m \in \mathbb{Z}_{N/{\rm gcd}(N,2)}$, $p\in \mathbb Z_{\frac{N-4}{k}}$, and $p'\in \mathbb Z_{\frac{N+4}{k}}$, then the faithful global symmetry is given by\footnote{Here, we are being sloppy to avoid clutter. What we write as $U(1)_A\times \mathbb Z_r$ must be understood as
\beq
\frac{U(1)_A \times \mathbb{Z}_{\scriptscriptstyle N_\psi + N_\chi}}{\mathbb{Z}_{\scriptscriptstyle ( N_\psi + N_\chi ) / r}}~.
\eeq
 }(we do not include the spacetime symmetry group)
\beq
 \frac{SU\left(\frac{N-4}{k}\right)_\psi \times SU\left(\frac{N+4}{k}\right)_\chi \times U(1)_A}{\mathbb{Z}_{N / {\rm gcd}(N,2)} \times \mathbb{Z}_{(N-4)/k} \times \mathbb{Z}_{ (N+4)/k}} \times \mathbb{Z}_r \times \mathbb{Z}_{\scriptscriptstyle {\rm gcd} (N,2)}^{(1)}~.
 \label{the global structure of the global symmetry}
\eeq
However, when not all values of $m,p,p'$ satisfy eqs. (\ref{eq:detailed cocycles psi}) and (\ref{eq:detailed cocycles chi}), the group we mod by is reduced to a subgroup of $\mathbb{Z}_{N / {\rm gcd}(N,2)} \times \mathbb{Z}_{(N-4)/k} \times \mathbb{Z}_{ (N+4)/k}$. As we discussed before, the discrete symmetry $\mathbb Z_r$ is also reduced to the genuine $\mathbb Z_p^{d\chi}\subset \mathbb Z_r$ discrete chiral symmetry in case that generators of $\mathbb Z_r$ are redundant to those of the color and flavor center groups,. This needs to be checked on a case by case basis.

\subsubsection{`t Hooft anomalies: old and new}
\label{`t Hooft anomalies: old and new}

`t Hooft anomalies are probed by performing global symmetry transformations in the presence of background gauge fields of the global symmetries, and are characterized by the non-trivial phases of the partition function ${\cal Z}$ under such transformations. While those phases vanish when background fields are turned off (hence the presence of those anomaly phases do not signal inconsistency of the theory) they nevertheless form an important set of consistency conditions on the IR phase. 
In addition to the standard `t Hooft anomalies such as $SU (n_\psi)^3$, $SU (n_\chi)^3$, $U(1)_A^3$, and so on, we acquire one completely new `t Hooft anomaly as well as several modified anomaly phases, thanks to the non-trivial center fluxes. We start with the standard `t Hooft anomalies.

\subsubsection*{(I) Perturbative $\left[SU (n_\psi)\right]^3$ and $\left[SU (n_\chi)\right]^3$ anomalies:}

These are the traditional 't Hooft anomalies associated to the $0$-form symmetries.  Anomaly inflow from $5\rightarrow 4$ dimensions, see, e.g., \cite{Callan:1984sa, Yonekura:2016wuc, Witten:2019bou}, provides a particularly elegant way to describe anomalies. See Appendix \ref{The descend procedure and counter terms} for a detailed account of the descend procedure and counter terms one can add in order to get rid of non genuine anomalies. 
Then,  the anomaly inflow is captured by 5$d$ Chern-Simons actions
\beq
\begin{split}
&\left[SU(n_\psi)\right]^3 : \; e^{i \int_{\mathbb M^5} \kappa_{\psi^3} \omega_5 (A^\psi)}~, 
\\[3pt]
& \left[SU(n_\chi)\right]^3 : \; e^{i \int_{\mathbb M^5} \kappa_{\chi^3} \omega_5 (A^\chi)}~,
\end{split}
\eeq
where the anomaly coefficients are 
\beq
\kappa_{\psi^3} = \mbox{dim}_\psi~, \quad \kappa_{\chi^3} = \mbox{dim}_\chi~.
\eeq
The 5$d$ Chern-Simons action $\omega_5 (A^f)$ (with $f=\psi,\chi$) is defined through the descent equation
\beq
d \omega_5 (A^f) = \frac{2\pi}{3 !} {\rm tr}_{\Box} \left( -\frac{i}{2\pi } \left(\hat{F}^f_2 - B_2^f \mathbf{1}\right) \right)^3\,,
\eeq 
where $\hat{F}^f_2$ is the $U(n_f)$ field strength and $B_2^f $ is the center background field, see Appendix~\ref{Descending in the ACF background fluxes} for details. The 4$d$ anomaly phase is obtained through the standard step: $\delta_\alpha \omega_5 (A^f) = d \omega_4^{(1)} (A^f, \alpha^f)$, where $\alpha^f$ is the gauge transformation parameter. We remind the reader that the anomaly and associated inflow action are defined up to local counter terms.

\subsubsection*{(II) Gravitational anomalies:}

 The mixed $U(1)_A$-gravitational and $\mathbb Z_p^{d\chi}$-gravitational  anomalies are described by the following anomaly inflow actions:
\beq
 U(1)_A [\mbox{grav}]^2 &:& \; e^{i \int_{\mathbb M^5} \frac{-\kappa_{ug}}{24} A_1 \wedge p_1 (\mathbb M^5)} \label{eq:Grav-U(1)_anomaly}~, 
 \\[3pt]
 \mathbb{Z}_p^{d\chi} [\mbox{grav}]^2 &:& \; e^{i \int_{\mathbb M^5}  \frac{-\kappa_{zg}}{24} z \wedge p_1 (\mathbb M^5)}~, \label{eq:Grav_Z_p_anomaly}
\eeq
where the anomaly coefficients are 
\beq
\kappa_{ug} = q_\psi \mbox{dim}_\psi n_\psi + q_\chi \mbox{dim}_\chi n_\chi~, \quad \kappa_{zg} = \mbox{dim}_\chi n_\chi~.
\eeq
$A_1$ and $z$ are $U(1)_A$  and $\mathbb{Z}_p^{d\chi} \subseteq \mathbb{Z}_r$ gauge fields, respectively, and $p_1 (\mathbb M^5) = - \frac{1}{8\pi^2} {\rm tr} (R\wedge R)$ is the first Pontryagin class of the tangent bundle where $R$ is the curvature $2$-form. One notes that $\int_{\mathbb M^4}p_1 (\mathbb M^4) \in 48\, \mathbb{Z}$ on a spin-manifold. So, there are two gravitational zero modes per Weyl fermion according to the index theorem and the Dirac index is ${\cal I}=2\kappa_{ug}$. It is easily seen that, under the $U(1)_A$ transformation $A_1 \to A_1 + d \alpha$, the anomaly inflow action eq.~(\ref{eq:Grav-U(1)_anomaly}) correctly reproduces the mixed gravitational--$U(1)_A$ anomaly on the spacetime manifold $\mathbb M^4 = \partial \mathbb M^5$. Likewise, the mixed gravitational--$\mathbb{Z}_p^{d\chi}$ anomaly is correctly captured under $z \to z + \frac{2\pi}{p} k_p$, $k_p \in \mathbb{Z}_p$. One way to think about the discrete symmetry in this case is to embed $\mathbb{Z}_p$ into a $U(1)$ group  and then restrict  the group parameters $z$ to a $\mathbb{Z}_p^{d\chi}$ subgroup of $U(1)$.

The existence/absence of the mixed gravitational-$\mathbb{Z}_p^{d\chi}$ anomaly plays an important role in the anomaly matching when $\mathbb{Z}_p^{d\chi}$  is not completely broken by a vacuum condensate. As we discuss in Sections~\ref{subsubsec:absence of composite fermions} and \ref{subsubsec:AM_unbroken_discrete_symm}, if the theory does not admit composite fermions and the vacuum leaves an unbroken subgroup of the  discrete chiral symmetry (with a non-vanishing mixed gravitational anomaly) on spin manifold, the vacuum condensate is ruled out. Similarly, if the condensate breaks $U(1)_A$ to a discrete subgroup $\mathbb Z_q \subset U(1)_A$, the anomaly  $\mathbb Z_q \left[\mbox{grav}\right]^2$ must vanish mod $q$. If this is not the case, the condensate is ruled out, see Section \ref{subsubsec:AM_unbroken_discrete_symm} for more detail.

\subsubsection*{(III) (0-form)-(1-form) mixed anomalies (CFU anomalies):}

These are new anomalies that we identify in this paper for the first time. The mixed  (0-form)-(1-form) anomalies in the CFU-background (we call them CFU-anomalies) are found by performing a global transformations of $U(1)_A \times \mathbb Z_{p}^{d\chi}$ in the background of the CFU fluxes. Non-trivial CFU anomalies show up as non-trivial phases of the partition function ${\cal Z}$ upon performing the corresponding global transformation:
\beq
U(1)_A \left[\mbox{CFU}\right]&:& {\cal Z}\rightarrow e^{i 2\pi \alpha\left[ q_\psi {\cal I}_{\psi} + q_\chi {\cal I}_{\chi}\right]}{\cal Z}~, \label{eq:CFU anomalies U(1)_R}
\\[3pt]
\mathbb Z_{p}^{d\chi}  \left[\mbox{CFU}\right]&:& {\cal Z}\rightarrow e^{i \frac{2\pi}{p}{\cal I}_{\chi}}{\cal Z}~,
\label{eq:CFU anomalies Z_p}
\eeq
where the Dirac indices are given by eqs. (\ref{eq:Dirac index psi}) and (\ref{eq:Dirac index chi}). 
While the expression of the Dirac index in the $U(1)_A \left[\mbox{CFU}\right]$ anomaly includes the contribution from $SU(N)$, it nonetheless vanishes  because the associated anomaly coefficient is identically zero (remember that $U(1)_A $ is a good symmetry in the background of $SU(N)$ fluxes). On the contrary, this is not true for the $\mathbb Z_{p}^{d\chi} \left[\mbox{CFU}\right]$-anomaly due to the discrete nature of the anomaly. 

The anomaly inflow action for the $U(1)_A \left[\mbox{CFU}\right]$-anomaly is (notice that the color fluxes do not contribute)
\beq \label{eq:inflow_action_U(1)}
\begin{split}
\exp i \int_{\mathbb M^5}  A_1 \wedge \Big [ & \sum_{f=\psi,\, \chi}\kappa_{uf^2} \left( c_2 (f) + \frac{n_f - 1}{2 n_f} w_2 (f) \wedge w_2 (f) \right) 
\\[3pt]
&+ \frac{\kappa_{u^3}}{24\pi^2} \left( F_2-B_2 \right) \wedge \left( F_2-B_2 \right) \Big ]~,
\end{split}
\eeq
where $A_1$ is the 1-form gauge field for $U(1)_A$, and $c_2 (f=\psi, \chi) = \frac{1}{8\pi^2} {\rm tr} \left( F_2^f \wedge F_2^f \right) \in \mathbb Z$ is the second Chern class of $SU(n_f)$. The anomaly coefficients in eq.~(\ref{eq:inflow_action_U(1)}) are given by
\beq
\kappa_{u\psi^2} = q_\psi \mbox{dim}_\psi~, \quad
\kappa_{u\chi^2} = q_\chi \mbox{dim}_\chi~, \quad 
\kappa_{u^3} = q_\psi^3 n_\psi \mbox{dim}_\psi  + q_\chi^3 n_\chi \mbox{dim}_\chi~.
\eeq
The sum of the $c_2$-term and the fractional instanton is nothing but the instanton number in the presence of the center background field. The last term in eq.~(\ref{eq:inflow_action_U(1)}) is the $U(1)_A$ contribution with its center background field $B_2$. 
It is straightforward to check that the $U(1)_A$ transformation $A_1 \to A_1 + d \alpha$ of the anomaly inflow action eq.~(\ref{eq:inflow_action_U(1)}) reproduces the anomaly eq.~(\ref{eq:CFU anomalies U(1)_R}) on the boundary $\partial \mathbb M^5 = \mathbb M^4$. One notes that the $U(1)_A \left[\mbox{CFU}\right]$ anomaly includes the traditional anomalies $U(1)_A \left[SU (n_\psi)\right]^2$, $U(1)_A \left[SU (n_\chi)\right]^2$, and $\left[U(1)_A\right]^3$ as a subset. This can be seen by switching off the fractional fluxes in eq.~(\ref{eq:inflow_action_U(1)}) and keeping only the second Chern Classes.

The $U(1)_A \left[\mbox{CFU}\right]$ anomaly plays an important role, similar to the mixed $U(1)_A\left[\text{grav}\right]^2$ anomaly, when we consider anomaly matching in the IR. If a given condensate breaks $U(1)_A$ to a nontrivial discrete subgroup $\mathbb Z_q \subset U(1)_A$ in a bosonic theory, the anomaly $\mathbb Z_q \left[\mbox{CFU}\right]$ must evaluate to $0$ mod $q$. If this is not the case, we argue in Section \ref{subsubsec:AM_unbroken_discrete_symm} that such condensate is ruled out. 
Similarly, the inflow action for the $\mathbb Z_{p}^{d\chi} \left[\mbox{CFU}\right]$ anomaly can be written as (contrary to the $U(1)_A \left[\mbox{CFU}\right]$-anomaly, the color fluxes do contribute here)
\beq
\begin{split}
&\exp i \int_{\mathbb M^5} z \wedge \Big [ 
\kappa_{zc^2} \left( c_2 (c) + \frac{N - 1}{2 N} w_2 (c) \wedge w_2 (c) \right)
\\[3pt]
& \quad + \sum_{f=\psi, \chi} \kappa_{zf^2} \left( c_2 (f)+ \frac{n_f - 1}{2 n_f} w_2 (f) \wedge w_2 (f) \right) +  \frac{\kappa_{zu^2}}{8\pi^2} \left( F_2-B_2 \right) \wedge \left( F_2-B_2 \right) \Big ]~, 
\end{split}
\label{eq:inflow_action_Z_p}
\eeq
where $z$ is the background gauge field of $\mathbb{Z}_p^{d\chi}$ and $c_2(c)$ is the second Chern class of $SU(N)$. The anomaly coefficients in eq.~(\ref{eq:inflow_action_Z_p}) are given by
\beq
\begin{split}
& \kappa_{zc^2}  = T_\psi n_\psi+T_\chi n_\chi~, \quad 
   \kappa_{z\chi^2} = \mbox{dim}_\chi~, \quad
   \kappa_{z\psi^2} = \mbox{dim}_\psi~,
\\[3pt] 
& \kappa_{zu^2} = q_\psi^2 n_\psi \mbox{dim}_\psi  + q_\chi^2 n_\chi \mbox{dim}_\chi~.\\
\end{split} 
\label{eq:anomaly_coeff_zc_zchi_zpsi_zu}
\eeq
It is understood that when $\mathbb{Z}_p^{d\chi}$ is represented to act only on $\chi$, we simply drop the $\psi$-contributions.
Under a $\mathbb{Z}_p^{d\chi}$ transformation $z \to z + \frac{2\pi}{p} k_p$, $k_p \in \mathbb{Z}_p$, the anomaly inflow action in eq.~(\ref{eq:inflow_action_Z_p}) generates the correct $\mathbb{Z}_p^{d\chi}$ anomaly (see eq. (\ref{eq:CFU anomalies Z_p}))  in the CFU background. In particular, this anomaly inflow action contains a new (0-form)-(1-form) mixed anomaly which is absent when no CFU-background is activated. Interestingly, in many cases, turning on only the $\mathbb Z_2$ color-center is not enough to guarantee a new anomaly, and it is crucial to activate more general CFU backgrounds, which yields a refined probe of the theory. We discuss this aspect in detail in Section~\ref{subsec:SU(8)_k4} and ~\ref{subsec:SU(8)_k2} with explicit examples.  
Again, the $\mathbb Z_{p}^{d\chi} \left[\mbox{CFU}\right]$ anomaly plays a pivotal role in ruling out vacuum condensates if they fail to break $\mathbb Z_p^{d\chi} $ completely.

\subsubsection*{(IV) Non-perturbative anomalies:}

Finally, let us briefly mention that in addition to all the above perturbative anomalies, the theory may also have nonperturbative ones. For example, if the theory admits an $SU(2)_\psi$ or $SU(2)_\chi$ global symmetries, it has a Witten anomaly if $\mbox{dim}_\psi$ or $\mbox{dim}_\chi$ is odd. However, a simple arguement can be invoked to show that there exists no Witten anomaly in bosonic chiral gauge theory. Explicitly, we remind that for bosonic theory we have $N = 4 \mathbb{Z}$ and from this it is quick to see that both $\mbox{dim}_\psi$ or $\mbox{dim}_\chi$ are even. This may also be seen by realizing that the fermion number $(-1)^F$ is gauged in bosonic theory and no fermion states are available to match Witten anomaly.

The theory may also possess $[\mathbb Z_p^{d\chi}]^3$ and $U(1)_A[\mathbb Z_p^{d\chi}]^2$ nonperturbative anomalies if the corresponding bordism group has a nonvanishing torsion. We do not discuss this type of anomalies in our work. 

\subsection{Theory on Nonspin Manifolds}
\label{Theory on Nonspin Manifolds}

In this section, we consider turning the CFU fluxes on nonspin manifolds, which may lead to additional constraints on the IR dynamics. Naively, a nonspin manifold does not admit fermions in the sense that there is an obstruction in lifting the $SO(4)$ bundle to a  $\mbox{Spin}(4)$ bundle on a 4$d$ manifold. On a nonspin manifold the Dirac index ${\cal I}=\frac{1}{192\pi^2}\int_{\mathbb M^4}\mbox{tr} \left[R\wedge R\right]$ is no longer an integer. However, since the signature of the manifold, which is defined as $\frac{1}{24\pi^2}\int_{\mathbb M^4}\mbox{tr} \left[R\wedge R\right]$, is always an integer, we can deduce that ${\cal I}\in  \frac{\mathbb Z}{8}$.  This point can also be seen by reminding that for an orientable vector bundle $\pi_{\mathbb E} : \mathbb E \to \mathbb M^4$ ($\mathbb E$ is the tangent bundle $\mathbb {TM}$ of $\mathbb M^4$), a spin structure exists on $\mathbb E$ if and only if the second Stiefel-Whitney class $w_2 (\mathbb M^4) \in H^2 (\mathbb M^4, \mathbb{Z}_2)$ vanishes. If $w_2 (\mathbb M^4)$ does not vanish, the triple overlap condition is not satisfied.

 In ${\rm Spin}_c$ manifold, a simple example of a nonspin manifold, this obstruction is remedied by exciting a $U(1)$ background flux $F_2$ with non-integral Chern class. Specifically, one demands $\int F_2\in \pi (2\mathbb Z+1)$ on 2-cycles, and thus, $\int_{\mathbb M^4}F_2\wedge F_2\in  \frac{\mathbb Z}{8}$. While the cocycle conditions are failed for individual ${\rm Spin} (4) / \mathbb{Z}_2 = SO(4)$ and $U(1)$ bundles, one nonetheless can obtain a consistent structure:
\beq
{\rm Spin}_c (4)= \frac{ {\rm Spin} (4) \times U(1) }{\mathbb{Z}_2}~,
\eeq 
where the quotient is with respect to the normal $\mathbb{Z}_2$. Simply put, the transition function is the product of ${\rm Spin} (4)$ and $U(1)$, and while each factor alone does not satisfy the triple overlap condition, the total transition function does: denoting $(\mathcal{G}, u) \in {\rm Spin} (4) \times U(1)$, we have
\beq
( \mathcal{G}, u )_{ij} \circ ( \mathcal{G}, u )_{jk} \circ ( \mathcal{G}, u )_{ki} = (-1, -1) \sim (1,1)~.
\label{cocylce on nonspin}
\eeq
Turning on the $F_2$ flux guarantees the integrality of the Dirac index.

A similar construction was discussed in the context of $SU(2)$ gauge theory with adjoint fermions in \cite{Witten:1995gf, Cordova:2018acb}, and there the relevant structure is described by:
\beq
\frac{{\rm Spin} (4) \times SU(2)}{\mathbb{Z}_2}~.
\eeq
The modding by $\mathbb{Z}_2$ implies that the second Stiefel-Whitney classes of the two factor groups are identified: $w_2 (\mathbb M^4) = w_2 (SU(2))$.

We shall activate general CFU-background fluxes that enable us to add fermions  on a nonspin manifold.  As we will see through the examples in Section~\ref{subsec:SU(8)_k2}, general CFU-backgrounds include the case of ${\rm Spin}_c$ structure when only $U(1)_A$ center flux is turned on, and it also includes the case with $\mathbb{Z}_2$ center flux of either the gauge group or non-abelian flavor group. The latter is a generalization of the situation appearing in adjoint $SU(2)$ theory mentioned above. We find that the most refined background fluxes with non-trivial additional `t Hooft anomaly in general require excitation of full color, non-abelian flavor, and $U(1)$ center fluxes.

The canonical example of a nonspin manifold is $\mathbb {CP}^2$.   It has a single $2$-cycle $\mathbb {CP}^1\subset \mathbb {CP}^2$ that supports  the $U(1)$ background flux. Putting fermions on $\mathbb {CP}^2$ in the background of CFU-fluxes was pursued  in \cite{Anber:2020gig} where we refer the reader for the details. Here, it is enough to state that we can use combinations of color, non-abelian flavor, and $U(1)_{A}$ center fluxes to render the fermions well-defined on $\mathbb {CP}^2$. 
We can repeat the discussion of Section \ref{Background Gauge Fields and Global Structure of the Symmetry Group}, and it leads to the following modified consistency equations:
\begin{eqnarray}
&& \psi \; : \;\; \underbrace{e^{i2\pi\frac{2m}{N}}}_{z_c} \underbrace{e^{i2\pi \frac{pk}{N-4}}}_{z_f} \underbrace{e^{-i 2\pi s \frac{(N+4)(N-2)}{kr}}}_{z_u} = -1 \label{eq:detailed cocycles psi nonspin}~, \\[3pt]
&& \chi : \;\;  \underbrace{e^{-i2\pi\frac{2m}{N}}}_{z_c} \underbrace{e^{-i2\pi \frac{p'k}{N+4}}}_{z_f} \underbrace{e^{i 2\pi s\frac{(N-4)(N+2)}{kr}}}_{z_u}=-1~,
\label{eq:detailed cocycles chi nonspin}
\end{eqnarray}
where $m \in \mathbb \mathbb{Z}_{N/{\rm gcd}(N,2)}$, $p\in \mathbb Z_{\frac{N-4}{k}}$, $p'\in \mathbb Z_{\frac{N+4}{k}}$ and $s$ is a $U(1)_A$ parameter. The minus sign on the right hand side compensates for  the minus sign arising from parallel transporting the spinor fields around appropriate closed paths in $\mathbb {CP}^2$ (see eq. (\ref{cocylce on nonspin}) and the discussion in \cite{Anber:2020gig}). Assuming that a solution to eqs.~(\ref{eq:detailed cocycles psi nonspin}) and (\ref{eq:detailed cocycles chi nonspin}) can be found, the topological charges corresponding to the CFU fluxes and gravity are given by~\footnote{The factor of $\frac{1}{2}$ is due to that the Pontryagin square is not an even integer on a nonspin manifold.}
\begin{eqnarray}
\begin{split}
Q_{c}&=\frac{m^2}{2}\left(1-\frac{1}{N}\right)~, \quad Q_{\psi}=\frac{p^2}{2}\left(1-\frac{k}{N-4}\right)~,
\\[3pt]
Q_{\chi}&=\frac{p'^2}{2}\left(1-\frac{k}{N+4}\right)~, \quad Q_{u}=\frac{1}{2}s^2\,, \quad Q_g=-\frac{1}{8}~.
\end{split}
\end{eqnarray}
Given the above topological charges, the Dirac-indices of both $\psi$ and $\chi$ are given by
\begin{eqnarray}
{\cal I}_{\psi}^{\mathbb {CP}^2}&=&n_\psi T_{\psi}Q_{c}+ \mbox{dim}_\psi  Q_{\psi}+\mbox{dim}_\psi n_\psi  \left(q_\psi^2 Q_{u}+Q_g\right)~,
\\[3pt]
{\cal I}_{\chi}^{\mathbb {CP}^2}&=&n_\chi T_{\chi}Q_{c}+ \mbox{dim}_{\chi}  Q_{\chi}+\mbox{dim}_{\chi} n_\chi  \left(q_\chi^2 Q_{u}+Q_g\right)~,
\label{nonspin Dirac indices}
\end{eqnarray}
which is always an integer, as can be easily checked for all the cases in this work. We use this information to calculate both $U(1)_A$ and  $\mathbb Z_p^{d\chi}$ anomalies in the CFU background: 
\beq
U(1)_A \left[\mbox{CFU}\right]_{\mathbb {CP}^2}&:& {\cal Z}\rightarrow e^{i 2\pi \alpha\left[ q_\psi {\cal I}_{\psi}^{\mathbb {CP}^2} + q_\chi {\cal I}_{\chi}^{\mathbb {CP}^2}\right]}{\cal Z}~, \label{eq:CFU anomalies U(1)_R CP2}
\\[3pt]
\mathbb Z_{p} \left[\mbox{CFU}\right]_{\mathbb {CP}^2}&:& {\cal Z}\rightarrow e^{i \frac{2\pi}{p}{\cal I}_{\chi}^{\mathbb {CP}^2}}{\cal Z}~.
\label{eq:CFU anomalies Z_p CP2}
\eeq
It is important to realize that we can have a theory with no CFU-anomaly on a spin manifold, but then the theory displays a non-trivial CFU-anomaly once  it is placed on a nonspin manifold. We present one such example in Section~\ref{subsec:SU(8)_k2}.

\section{Matching the Anomalies in the IR}
\label{Matching the Anomalies in the IR}

In this section, we discuss matching of `t Hooft anomalies in the IR \cite{tHooft:1979rat}. Our strategy is to assume that the theory confines~\footnote{It follows that when $N$ is even, the fundamental Wilson line operator obeys the area law,  or equivalently, the 1-form  $\mathbb Z_2^{(1)}$ center symmetry is  unbroken.} and examine whether all UV `t Hooft anomalies are matched in the IR.\footnote{In this work we mainly test if a theory can confine in the IR satisfying all the anomaly matching conditions, and assume it flows to an IR CFT as the most plausible possibility when the theory is unlikely to confine. However, we note that it is nevertheless possible for the theory to flow to a more exotic phase such as a Coulomb phase. In this case, the IR phase may be described by some IR free gauge group with fermions charged under it, and these fermions may saturate the anomalies seen in UV theory. We leave this possibility for future investigation.} To this end, the standard method is to survey possible vacuum condensates and check their symmetry breaking pattern, mostly focusing only on a vacuum operator with the smallest scaling dimension. However, we show that the combination of all anomalies, continuous vs discrete, and their matching  enable us to check the entire space of candidate vacuum operators. In Section~\ref{subsec:SU(8)_k4}, this procedure  eliminates many condensates and may point out to the most-likely IR scenario. For the theory discussed in Section~\ref{subsec:SU(8)_k2}, our procedure suggests that it is likely that the theory flows to an IR CFT.

The symmetry breaking pattern $G \to H \subset G$ is determined by the existence/absence of allowed gapless modes and/or TQFTs that match the UV anomalies. 

A UV theory can have both continuous and discrete symmetries $G = G_c \times \mathbb Z_p^{d\chi}$ (here we are not concerned with the global structure of the symmetry discussed around eq. (\ref{the global structure of the global symmetry})), which are spontaneously broken down  to $H = H_c \times H_d$ by a given vacuum condensate. One notes that $H_d$ is not necessarily a subgroup of $\mathbb Z_p^{d\chi}$, for instance, $U(1)_A$ may as well be broken to a discrete subgroup. 

The following general remarks are in order. 

\begin{itemize}
\item[(i)] If $H_c$ has non-trivial `t Hooft anomalies, they can be matched by composite fermions \cite{Frishman:1980dq,Coleman:1982yg}. If the theory does not admit composite fermions, as in the case  $N = 4 \mathbb{Z}$ (see Section~\ref{subsubsec:absence of composite fermions}), these anomalies cannot be matched in a confining theory. 
\item[(ii)] If $H_d$ is anomalous, those anomalies must be matched by either massless fermions or  $H_d$-preserving TQFT.  In the absence of composite fermions, the only means of achieving anomaly matching is via a TQFT. One may use the results of \cite{Witten:2016cio, Wang:2017loc, Tachikawa:2017gyf, Wan:2018djl, Cordova:2019bsd, Cordova:2019jqi} to check whether a symmetry preserving and unitary TQFT may exist.  
\item[(iii)] In theories with no composite fermions, a combination of (i) and (ii) above puts a strong constraint on the IR phase. If no vacuum operators exist such that $H_c$ is anomaly-free and either $H_d = \emptyset$ or a TQFT exists, we conclude that the theory cannot confine and generate a mass gap. In this situation, the most reasonable candidate for the IR phase is an (interacting) conformal field theory. 
\item[(iv)] The anomaly associated with the broken part of $G_c$ is matched by a Wess-Zumino-Witten (WZW) term of Goldstone bosons. It will be discussed in Section~\ref{subsubsec:AM_broken_continuous_symm}.
\item[(v)] For anomalies that involve the discrete chiral symmetry $\mathbb Z_p^{d\chi}$, the matching can be achieved by the full breaking of $\mathbb Z_p^{d\chi}$ (and the formation of domain walls), a symmetry-preserving  TQFT, or a partial breaking of $\mathbb Z_p^{d\chi}$ accompanied by a TQFT. The matching of these anomalies by domain walls is discussed in Section~\ref{subsubsec:AM_broken_discrete_symm}. 

\end{itemize}

\subsection{Absence of composite fermions}
\label{subsubsec:absence of composite fermions}

Interestingly, most of chiral gauge theories considered in this work do not admit gauge invariant composite fermion operators.  We show below that composite fermion operators in the 2-index chiral gauge theory are prohibited when $N = 4 \ell$ $(\ell \in \mathbb{Z})$. Combined with the obstruction of a symmetry-preserving TQFT, this leads to a severe limitation on the confinement  as an allowed IR phase of the theory. 

Any gauge-invariant composite fermion operator can be written as 
\beq
\mathcal{O}_\text{fermion} \sim \epsilon^{a_1} \psi^{a_2} \chi^{a_3} (f^c)^{a_4}~,
\eeq
where $\epsilon$ is the antisymmetric Levi-Civita symbol. $\psi$ carries two upper indices and $\chi$ carries $(N-2)$ upper indices,  while $f^c$ represents the color field strength with one upper and one lower indices. The powers $a_1, a_2, a_3, a_4$ denote the number of insertions of the corresponding object. In general, we need to insert powers of the color field $f^c$ to guarantee that $\mathcal{O}_\text{fermion} $ does not vanish because of symmetry reasons. Demanding that $\mathcal{O}_\text{fermion}$ is a color singlet and fermionic operator give the two conditions  $2 a_2 + (N-2) a_3= a_1 N$ and $a_2 + a_3 = 2k +1$ $(k \in \mathbb{Z})$, respectively. Notice that   $f^c$ carries one upper and one lower color indices, and thus,  insertions of $f^c$  in $\mathcal{O}_\text{fermion}$ does not affect the counting of  indices and $a_4$ does not appear in the two conditions. Combing the conditions give
\beq \label{eq:noferminoop:combined}
(N-4) a_3 - a_1 N + 4k + 2 = 0~.
\eeq
For $N = 4 \ell$ $(\ell \in \mathbb{Z})$, eq.~(\ref{eq:noferminoop:combined}) reduces to $2 \left( \ell a_1 +(1- \ell) a_3  - k \right) = 1$, which has no solution. Therefore,  2-index $SU(4 \ell)$ chiral gauge theories do not admit composite fermions. This includes all but $N=5,6,$ and 10 cases in Table (\ref{tab:2-index_chiral_theory}).

There is a simple alternative explanation to the above finding. We first recall that the theory has the 0-form color center $\mathbb{Z}_{N / {\rm gcd} (N,2)}$. Then, it is sufficient to realize that for $N=4\ell, \ell \in \mathbb{Z}$, the 0-form color center symmetry is $\mathbb{Z}_{2\ell}$ which includes $\mathbb{Z}_2^{\scriptscriptstyle \rm L}$, the fermion number $(-1)^F$ of the Poincar\'e group, as a subgroup. In other words, this particular choice of the gauge group and fermion representations has an effect of gauging the fermion number, resulting in every gauge-invariant operator being bosonic. To show that this is the only choice of $N$ that leads to a bosonic theory, we simply note that for $N = 2\ell +1$ and $N=4\ell +2$, the 0-form color center is $\mathbb{Z}_{2\ell + 1}$ and so the fermion number is left ungauged.

\subsection{Unbroken discrete symmetry: TQFTs and obstructions}
\label{subsubsec:AM_unbroken_discrete_symm}

In principle, anomalies of unbroken discrete symmetries $H_d$, either from $G_c$ or $\mathbb Z_p^{d\chi}$, can be matched by composite fermions or symmetry-preserving TQFTs. However, in a theory with no composite fermions, these anomalies must be saturated by a TQFT. 
A criteria for the obstruction of such TQFTs has been proposed by the authors of \cite{Cordova:2019bsd,Cordova:2019jqi}. Here, we give a succinct summary of these results. 

\subsubsection*{Obstruction from mixed gravitational anomalies}

We consider a mixed anomaly of a discrete symmetry $\mathbb{Z}_n$ (this can be unbroken subgroup of $U(1)_A$ or $\mathbb{Z}_p^{d\chi}$) and Poincar{\'e} symmetry. Assuming that the theory on a compact smooth spin 4-manifold $\mathbb M^4$ has $k$ fermions charged under $\mathbb{Z}_n$, there are $2k$ fermion zero modes if the Pontryagin number $\int_{\mathbb M^4} p_1 (\mathbb {M}^4)$ is 48 (this is the minimal number of zero modes in the background of a spin gravitational instanton). The $\mathbb{Z}_n$ anomaly is then described by a $(4+1)$d anomaly inflow action, or SPT (symmetry-protected topological order). It is known that if $n$ is a divisor of $2k$ (the number of fermion zero modes in gravitational instanton background), then the SPT admits a symmetry-preserving gapped boundary. On the other hand, the authors of \cite{Cordova:2019jqi}  showed that if $n$ is not a divisor of $2k$, then no such TQFT exists. The argument of \cite{Cordova:2019jqi} is based on the observation that while a symmetry-preserving boundary theory $\mathcal{B}$ implies a vanishing partition function $\mathcal{Z}_{\mathcal{B}} [\mathbb M^4] = 0$,  the partition function of any unitary spin TQFT on a simply connected smooth spin 4-manifold with a unique state on $S^3$ cannot be zero. 

In short, given  $\mathbb{Z}_n$ with a non-trivial gravitational anomaly we check the condition $n | 2k$. If not satisfied, then a symmetry-preserving TQFT is excluded.

\subsubsection*{Obstruction from mixed anomalies in the CFU background}

Next, we consider the mixed anomaly of a discrete symmetry $\mathbb{Z}_n$ (this can be unbroken subgroup of $U(1)_A$ or $\mathbb{Z}_p^{d\chi}$) and some other internal symmetry. In particular, we are interested in the mixed anomaly between 0-form discrete $\mathbb{Z}_n$ and 1-form symmetry, i.e.~CFU backgrounds. The corresponding anomaly inflow action is schematically given by 
\beq
\int \omega_5 (A) \sim \int z \wedge w_2 \wedge w_2
\eeq
where $z$ represents the $\mathbb{Z}_n$ discrete gauge field and the part $w_2 \wedge w_2$ represents the CFU fluxes. 
If the theory does not admit gauge invariant fermionic operators, the only hope for anomaly matching is via a $\mathbb{Z}_n$-preserving TQFT. The existence of such TQFT has been investigated in  \cite{Witten:2016cio, Wang:2017loc, Tachikawa:2017gyf, Wan:2018djl, Cordova:2019bsd, Cordova:2019jqi}. We follow the criteria of \cite{Cordova:2019bsd} to judge the presence/absence of such TQFT.  We put the theory  on the spin manifold $\mathbb M^4 = \mathbb S^{2} \times  \mathbb S^{2}$  and turn on the 1-form background fluxes, which pierce both spheres. If the anomaly inflow action evaluated on the mapping torus manifold $\mathbb N^5\equiv \mathbb S^1 \times \mathbb M^4$ is non-trivial, that is,
\beq
e^{2\pi i \int_{\mathbb N^{5}} \omega_{5} (A) } \neq 1
\eeq 
there exists no unitary symmetry-preserving TQFT on $\mathbb M^4$. We immediately conclude that a nontrivial $\mathbb Z_n$ anomaly in the CFU background cannot be matched by a spin TQFT.

\subsection{Broken continuous symmetries: the Wess-Zumnio-Witten term}
\label{subsubsec:AM_broken_continuous_symm}

We discuss the anomaly matching of the spontaneously broken continuous symmetry via goldstone bosons emphasizing that the Wess-Zumnio-Witten (WZW) term will always match anomalies for the broken continuous symmetry even after turning on the fractional 't Hooft fluxes. 

Consider a continuous symmetry $G$ broken down to an anomaly-free group $H$. We write the gauge field for $G$ as $A$ and gauge field for $H$ as $A_h$. Non-abelian anomaly satisfying the Wess-Zumino consistency condition is obtained via Stora-Zumino descent procedure and the WZW action is better understood in this language (see~\cite{Lee:2020ojw,Yonekura:2020upo} for a modern perspective). 
It takes the form
\begin{eqnarray}
S_{\scriptsize\mbox{WZW}}=\int_{\mathbb M^5}\left( \tilde{\omega}_5^{(0)} (A_h,A) - \tilde{\omega}_5^{(0)} ((A^{U^{-1}})_h, A^{U^{-1}}) \right)\,,
\label{eq:WZW}
\end{eqnarray}
where the shifted Chern-Simons action $\tilde{\omega}_5^{(0)} (A_h, A)$ is given by
\beq
\tilde{\omega}_{5}^{(0)} (A_h, A) =  \int_0^1 d s \; {\rm tr} \left[ (A - A_h) F_s^2 \right], \;\; F_s = d A_s + A_s^2, \;\; A_s = s A + (1-s) A_h~. \label{eq:shifted_CS}
\eeq
One nice property of the shifted Chern-Simons action $\tilde{\omega}_{5}^{(0)}$ is that it is invariant when the two arguments $A$ and $A_h$ transform the same way. It implies that $\tilde{\omega}_5^{(0)} (A_h, A)$ is invariant under any $H \subset G$ transformations provided the Lie algebra is reductive and symmetric~\footnote{Writing the algebra for $H$ as $\mathbf{h}$ and that for coset $G/H$ as $\mathbf{k}$, the Lie algebra is reductive if $[h, k ] \subset \mathbf{k}$ and further is symmetric if $[k , k] \subset \mathbf{h}$, where $h \in \mathbf{h}$ and $k \in \mathbf{k}$.}. If $H$ is anomaly-free, the shifted Chern-Simons action is just a modification of the canonical CS action by a local counter term in such a way that $H$ is free of mixed anomalies with $G$ (see~\cite{Hong:2020bvq} for a review).
$U$ in the WZW action in eq.~(\ref{eq:WZW}) is the Goldstone field parametrizing the coset $G/H$ and it is understood that a proper extension to $\mathbb M^5$ is perfomed. 
For the spacetime manifold $\mathbb M^4 = \mathbb S^4$, the 5$d$ manifold is taken to be 5-disk $\mathbb D^5$, but we consider a more general $\mathbb M^4$ and associated $\mathbb M^5$ such that $\partial \mathbb M^5 = \mathbb  M^4$. Also, $A^{U^{-1}}$ is the Maurer-Cartan $1$-form: 
\begin{eqnarray}
A^{U^{-1}} \equiv  U^{-1} A\, U +  U^{-1} d  U\,,
\end{eqnarray}
which satisfies the Maurer-Cartan structure equation
\begin{eqnarray}
d  A^{U^{-1}} +  A^{U^{-1}} \wedge  A^{U^{-1}} =  U^{-1}  F  U \,.
\end{eqnarray}
These are just gauge transformation of $A$ by $U^{-1}$ and its field strength $F$.
By construction, the WZW action satisfies the following three conditions which are necessary to guarantee that the Goldstone bosons match the UV anomaly: (1) it solves the anomalous ward identity, (2) it vanishes as $ U \rightarrow 1$, and thus,  background gauge fields cannot solely solve the anomalous ward identity, and (3) it is a function  of  the boundary values of $A$ and $ U$, i.e., it manifestly depends only on fields defined on $\mathbb M^4$ and does not depend on the extension to $5$ dimensions. The latter condition is easily seen by noting that $d \big ( \tilde{\omega}_5^{(0)} ({A^{U^{-1}}})- \tilde{\omega}_5^{(0)} (A) \big ) = \Omega_6 (A^{U^{-1}}) - \Omega_6 (A) = 0$ by the gauge invariance of the 6$d$ anomaly polynomial $\Omega_6$. This last property is what makes the WZW action an intrinsically 4$d$ local action~\footnote{Due to the closedness, (at least locally) $\big ( \tilde{\omega}_5^{(0)} ({A^{U^{-1}}})- \tilde{\omega}_5^{(0)} (A) \big ) = d \Gamma_4$ and we get $S_{\scriptstyle \rm WZW} = \int_{\mathbb M^4} \Gamma_4$. In fact, $S_{\scriptstyle \rm WZW}$ is independent of the way the extension $\mathbb{M}^4 \to \mathbb{M}^5$ is made provided it satisfies a quantization condition~\cite{Witten:1983tw}.}, although its symmetry property manifests better in its 5$d$ representation.

Regarding the anomaly matching, it is sufficient to mention that $A \rightarrow g A g^{-1} +g dg^{-1}$ and $U \rightarrow g\,  U^g h (U, g)$ under  $g \in G$ transformation. From this, it can be shown that, under a \emph{global} $g \in G$, $A^{U^{-1}}$ transforms as
\beq
A^{U^{-1}} \to h (U,g) A^{U^{-1}} h^{-1} (U,g) + h (U, g) d h^{-1} (U,g)~,
\eeq 
that is, $A^{U^{-1}}$ transforms as a particular \emph{local} gauge transformation (note that $U$ has spacetime dependence), and in particular as if it is a gauge connection of $H$. This fact can be combined with the invariance of the shifted CS action under simultaneous transformation of the two arguments to conclude that the second term in eq.~(\ref{eq:WZW}) is invariant under any $G$ transformation. Recalling that the first term in eq.~(\ref{eq:WZW}) is just the anomaly inflow action, it matches the anomaly of the UV theory. The first term in eq.~(\ref{eq:WZW}) is also invariant under $H$ transformation, hence is $S_{\scriptstyle \rm WZW}$. This implies that the WZW action does not capture any anomaly associated with the unbroken $H$. If the anomaly of $H$ does not vanish in the UV, it needs to be matched by massless composite fermions or TQFT (for discrete subgroups). $H$ can be a semi-simple or a direct product of semi-simple groups, continuous or discrete.

The same procedure can be repeated  in the presence of the CFU fluxes. The most general field strength including center background fields of color ($c$), flavor ($f$) and $U(1)$ groups is written as 
\beq
{\cal F}_2 = (\hat{f}^{c}_2 - B^{c}_2 \mathbf{1}_{c,N}) \otimes \mathbf{1}_f  + \mathbf{1}_c \otimes (\hat{F}^{f}_2 - B^{f}_2 \mathbf{1}_f)  + \mathbf{1}_c \otimes \mathbf{1}_f \otimes (F_2 - B_2)~,
\label{eq:general_F}
\eeq
where the field strength $\hat F^f$ is valued in $U(n_f)$, etc. $ \mathbf{1}_f $, $ \mathbf{1}_{c}$, $\mathbf{1}_{c,N}$ are unit matrices of dimensions $n_f$, $\mbox{dim}_{\cal R}$, and $N$, respectively.  See Appendix \ref{Descending in the ACF background fluxes} for details. 
The associated gauge fields may be defined in the standard manner. 
Starting with the 6d anomaly polynomial 
\beq
\Omega_6 (A) = \frac{2\pi}{3!}\sum_{\scriptsize\mbox{Weyl}}\mbox{tr}_{\cal R}\left(\frac{-i {\cal F}_2}{2\pi}\right)^3~,
\eeq
virtually identical steps can be taken to arrive at anomaly inflow actions for various anomalies, including 0-form and 1-form mixed anomaly, and the WZW action for continuous $G \to H$ (now including fractional center fluxes) can be constructed. We remark that 
in the discussion of spontaneously broken \emph{continuous} symmetry, the color-part (the first term in eq.~(\ref{eq:general_F})) can be dropped since its anomaly coefficient vanishes identically. Thus, the first term of the WZW action (\ref{eq:WZW}) will generate the  $U(1)_A$ anomaly in the CFU background flux, while the second term stays invariant per the discussion above; the full WZW action reproduces the $U(1)_A[{\rm CFU}]$ anomaly (when $U(1)_A$ is broken). In particular, we note that when $U(1)_A$ is broken down to a discrete subgroup $\mathbb Z_q\subset U(1)_A$,  we need to check that the  $\mathbb Z_q$ anomaly in the CFU background flux vanishes in the UV. If this were not the case, then the IR phase has to match this anomaly either by composite fermions or a TQFT. A detailed  discussion of the descend procedure in the absence/presence of CFU fluxes,  anomaly inflow action,  and  issue of counter terms is provided in Appendices \ref{The descend procedure and counter terms} and \ref{Descending in the ACF background fluxes}.

Finally, we also mention that we can write down a WZW action that matches the $U(1)_A[\text{grav}]^2$ anomaly by simply replacing $\tilde \omega_5^{(0)}$ by a mixed Chern-Simons term $A^1\wedge p_1 (\mathbb M^5)$, where $A_1$ is the $U(1)_A$ gauge field:
\begin{eqnarray}
- \frac{1}{24} \int_{\mathbb M^5}\left(A_1-A_1^{U^{-1}}\right) \wedge   p_1 (\mathbb M^5) ~,
\end{eqnarray}
and $A_1^{U^{-1}}$ is the ``gauge transformation" of $A_1$ by the goldstone field.

\subsection{Broken discrete symmetries: TQFT and domain walls}
\label{subsubsec:AM_broken_discrete_symm}

For broken $\mathbb Z_p^{d\chi}$, the $\mathbb Z_p^{d\chi}$ anomaly in CFU background can be matched by a TQFT that describes multiple vacua separated by domain walls~\cite{Gaiotto:2014kfa, Kapustin:2014gua,Anber:2020xfk,Anber:2021lzb}.   We first discuss a universal TQFT that matches the anomaly of broken $q$-form $\mathbb{Z}_n^{(q)}$. We then specialize to the mixed anomaly between 0-form $\mathbb{Z}_p^{d\chi}$ and CFU background. The final result eq. (\ref{eq:TQFT_DW}) matches those in~\cite{Anber:2020xfk,Anber:2021lzb}.

Consider a $\mathbb{Z}_n$ gauge theory in $d$ spacetime dimension described by a BF theory.
\beq
S = \frac{i n}{2\pi} \int a_q \wedge d b_{d-q-1}~.
\eeq
In addition to $(q-1)$-form and $(d-q-2)$-form gauge symmetries, the theory has two higher-form global symmetries which are $q$-form and $(d-q-1)$-form $\mathbb{Z}_n$ symmetries:
\beq
&& \mathbb{Z}_n^{(q)} : \; a_q \to a_q + \frac{1}{n} \epsilon_q \\
&& \mathbb{Z}_n^{(d-q-1)} : \; b_{d-q-1} \to b_{d-q-1} + \frac{1}{n} \epsilon_{d-q-1}~,
\eeq
where $\epsilon_q$ and $\epsilon_{d-q-1}$ are closed and have integral periods: $\int \epsilon_q, \int \epsilon_{d-q-1} \in 2\pi \mathbb{Z}$. These two higher-form symmetries have mixed anomalies. This may be seen by first coupling the theory to background gauge fields $A_{d-q}$ and $B_{q+1}$ of these global symmetries.
\beq
S \to \tilde{S} = \frac{i n}{2\pi} \int a_q \wedge \left( d b_{d-q-1} - A_{d-q} \right) - B_{q+1} \wedge b_{d-q-1}~.
\eeq
We have combined two terms to make the theory manifestly invariant under $\mathbb{Z}_n^{(d-q-1)}$. It also shows clearly that the relevant symmetry is non-linearly realized. Note that we could have exchanged the role of the two and make instead $\mathbb{Z}_n^{(q)}$-invariance (and its non-linear realization) manifest. Under $\mathbb{Z}_n^{(q)}$ transformation, however, $\tilde{S}$ shifts as
\beq
\tilde{S} \to \tilde{S} + \frac{i}{2\pi} \int \epsilon_q \wedge \left( d b_{d-q-1} - A_{d-q} \right)~.
\eeq
While the first term is consistent (note that it is $2\pi \mathbb{Z}$-valued), the second non-invariant term cannot be removed by any local counter terms, showing that the theory has a mixed anomaly:
\beq
\mathbb{Z}_n^{(q)} : \; \delta  \tilde{S} = - \frac{i}{2\pi} \int \epsilon_q \wedge A_{d-q}~. 
\eeq
This mixed anomaly may be described by an anomaly inflow action
\beq
S_{\scriptscriptstyle \rm inflow} = \frac{i n}{2\pi} \int A_{d-q} \wedge B_{q+1}~.
\eeq
The TQFT $\tilde{S}$ matches the anomaly of the spontaneously broken $\mathbb{Z}_n^{(q)}$.

Now we specialize to our case by setting  $q=0$, $d=4$, $n=p$, and writing $a_0 = \phi$:
\beq
S = \frac{i p}{2\pi} \int_{\mathbb{M}^4} \phi d b_3 - \frac{i p}{2\pi} \int_{\mathbb{M}^4} \phi A_4 - \frac{i p}{2\pi} \int_{\mathbb{M}^4} b_3 \wedge B_1~,
\label{eq:TQFT_DW}
\eeq 
with associated anomaly inflow action
\beq
S_{\scriptscriptstyle \rm inflow} = \frac{i p}{2\pi} \int_{\mathbb{M}^5} B_1 \wedge A_4~,
\eeq
where $B_1 = z$ is the $\mathbb {Z}_p^{d\chi}$ gauge field and $A_4$ is the CFU background fluxes. The exact matching can be obtained by comparing eqs.~(\ref{eq:inflow_action_Z_p}) and (\ref{eq:TQFT_DW}). We obtain
\beq
\int \frac{p}{2\pi} A_4 = \kappa_{zc^2} Q_c + \kappa_{z\chi^2} Q_\chi + \kappa_{z\psi^2} Q_\psi + \kappa_{zu^2} Q_u~, 
\label{eq:TQFT_DW_A_4_CFU}
\eeq
where $Q_\psi$ needs to be dropped if we choose  $\mathbb Z_{p}^{d\chi}$  to act  only on $\chi$.

The first term in eq.~(\ref{eq:TQFT_DW}) describes domain wall of broken $\mathbb{Z}_p^{d\chi}$. Specifically, $\langle e^{i \phi} \rangle$ is the order parameter for the $\mathbb{Z}_p^{d\chi}$ breaking: $\langle e^{i \phi} \rangle = e^{\frac{i 2\pi k}{p}}, \; k= 0, \cdots (p-1)$ labels the degenerate vacua. Also, $e^{i \oint b_3}$ describes domain walls separating those vacua. The second term then reproduces anomaly of the UV theory.

\section{Confinement with Chiral Symmetry Breaking }
\label{subsec:SU(8)_k4}

In this section, we analyze the bosonic chiral gauge theory  $SU(8)$ with $k=4$ and discuss its possible IR phases in detail. 
The theory may confine through the dimension-five fermion bilinear vacuum operator supplemented with another vacuum condensate e.g.~${\cal O} \sim \chi^4$, and its IR phase is described by a $(SU(3) \times U(1))/(\mathbb{Z}_4 \times \mathbb{Z}_3) \to SU(2)$ non-linear sigma model and $\mathbb Z_2^{d\chi}$-breaking domain walls. If, however, vacuum operators decouple along the RG trajectory, the theory will most likely run into an IR fixed point.

\subsection{$SU(8)$ with $k=4$ ($n_\psi = 1$, $n_\chi = 3$) }

The symmetries of this theory and the charges of the fermions are summarized in the following table:
 \begin{equation}\label{tab:charges su8 with k4}
\begin{tabular}{|c|c|c|c||c|c|c|}
\hline
& $SU(3)_{\chi}$ & $U(1)_{A}$ & $\mathbb Z_{2}^{d\chi}$ & $\mathbb{Z}_4^c$ & $\mathbb{Z}_3^\chi$ & $\mathbb{Z}_2^{\scriptscriptstyle \rm L}$ \\
 \hline
$\psi$ & $1$ & $-9$ & $0$ & $1$ & $0$ & $1$\\ 
\hline
$\chi$ & $\overline\Box$ & $5$ &$1$ & $-1$ & $-1$ & $1$ \\ 
\hline
\end{tabular}
\end{equation}
The theory enjoys a genuine discrete chiral symmetry $\mathbb Z^{d\chi}_{r=\mbox{gcd}(18,10)=2}$, which cannot be absorbed in the color or flavor centers. This symmetry acts on fermions as $(\psi,\chi)\rightarrow (\psi, -\chi)$, which is equivalent to $(\psi,\chi)\rightarrow (-\psi, \chi)$ using a $\mathbb{Z}_2^{\scriptscriptstyle \rm L}$ transformation~\footnote{ A $\mathbb{Z}_2^{\scriptscriptstyle \rm L}$ transformation acts on the fermions as $(\psi,\chi)\rightarrow (-\psi,-\chi)$.}. In order to determine the faithful global symmetry, we solve the consistency equations in eqs.~(\ref{eq:detailed cocycles psi}) and (\ref{eq:detailed cocycles chi}). There exists a solution for every $m\in \mathbb Z_4^c$ and $p' \in \mathbb Z_3^\chi$, and hence, the solutions form the group $\mathbb Z_4\times\mathbb Z_3$. Following the discussion around eq. (\ref{the global structure of the global symmetry}), the faithful global symmetry is
\begin{eqnarray}
\frac{SU(3)_\chi\times U(1)_{A}}{\mathbb Z_4\times \mathbb Z_{3}}\times \mathbb Z_2^{d\chi}\times \mathbb Z_2^{(1)}~,
\end{eqnarray}
and the 1-form center symmetry is understood to act on fundamental Wilson lines.

\subsection{Anomalies on spin manifold}

Here, we list all the anomalies on spin manifold, including the new CFU anomalies.  The fact that there exists multiple solutions to the consistency equations (\ref{eq:detailed cocycles psi}) and (\ref{eq:detailed cocycles chi}) means that we can turn on CFU fluxes, which will be used to find 't Hooft anomalies. It is instructive to consider three distinct solutions corresponding to three distinct fluxes.   They are given by 
\beq
(m, p', s) = (4,0,0)~, \quad (1,0,1/4)~, \quad (1,1,-1/12~)~.
\eeq
Notice that these solutions are not independent, and only two of them yield a constraint on the IR phase, as we shall explain. Yet, we choose to consider all three of them for pedagogical reasons. 
The first solution corresponds to activating only the $\mathbb{Z}_2^{(1)}$ color center, while the second corresponds to activating both $\mathbb{Z}_4^{(1)}$ color center together with the  $\mathbb{Z}_4^{(1)}$ subgroup of $U(1)^{(1)}$ center. Finally, the last one is turning on all three fluxes: $\mathbb{Z}_4^{(1)}$ color flux, $\mathbb{Z}_3^{(1)}$ flavor center, and $\mathbb{Z}_{12}^{(1)} \subset U(1)^{(1)}$ fluxes\footnote{\label{explaining U(1) background}$U(1)^{(1)}$ is the one-form global symmetry that is associated to gauging $U(1)_A$. Indeed, we cannot gauge the full $U(1)_A$ since it is anomalous. However, a discrete subgroup of it can be gauged, which gives rise to a $1$-form global symmetry.  This is guaranteed to be the case if the background flux of $U(1)_A$ that we excite satisfies the consistency conditions (\ref{eq:detailed cocycles psi}) and (\ref{eq:detailed cocycles chi}). In the current example, we see that there is a $\mathbb Z_{12}^{(1)}$ 1-form global symmetry associated with turning on a $\mathbb Z_{12}\subset U(1)_A$ flux. In fact, one can construct a Wilson line $W_{1}$ that is charged under the $\mathbb Z_{12}^{(1)}$ 1-form symmetry, such that $W_1\rightarrow e^{-i\frac{2\pi}{12}} W_1$ under this symmetry.  $W_{1}$ is not a genuine line operator (topological) on its own. However, the combination of  (i) $W_1$, (ii) the color Wilson line $W_c$, which transforms as $W_c\rightarrow e^{i\frac{2\pi}{4}} W_c$, and (iii) the flavor Wilson line $W_\chi$, which transforms as  $W_\chi\rightarrow e^{i\frac{2\pi}{3}} W_\chi$ is topological. In other words, the Wilson line ${\cal W}$ defined as ${\cal W}\equiv W_1 W_cW_\chi$ cannot be screened by a dynamical fermion, and thus, ${\cal W}$ is topological. }.   Below, we summarize the topological charges and Dirac indices (setting $n_1=n_2=0$) for each solution.
\begin{equation}\label{tab:top_charges_Dirac_index_8_k4}
\begin{tabular}{|c|c|c|c|}
\hline
& $(4,0,0)$ & $(1,0,\frac{1}{4})$ & $(1,1,-\frac{1}{12})$ \\
 \hline
$Q_c$ & $14$ & $\frac{7}{8}$ & $\frac{7}{8}$ \\ 
\hline
$Q_\chi$ & $0$ & $0$ & $\frac{2}{3}$ \\ 
\hline
$Q_u$ & $n_1 n_2$ & $\left( -\frac{1}{4} + n_1 \right) \left( -\frac{1}{4} + n_2 \right)$ &  $\left( \frac{1}{12} + n_1 \right) \left( \frac{1}{12} + n_2 \right)$  \\ 
\hline \hline
$\mathcal{I}_\psi$ & $140$ & $191$ & $29$ \\
\hline
$\mathcal{I}_\chi$ & $252$ & $147$ & $49$ \\
\hline
\end{tabular}~.
\end{equation}

The UV theory has the following `t Hooft anomalies:
\begin{itemize}
\item $\left[SU (3)_\chi\right]^3$ anomaly:  $\kappa_{\chi^3} = 28$. Unless the vacuum  breaks $SU(3)_\chi$ to an anomaly-free subgroup, essentially there is obstruction for the theory to confine in the IR. 
\item mixed gravitational anomalies: 
\begin{itemize}
\item mixed $U(1)_A$-gravity anomaly : $\kappa_{ug} = 96$. When $U(1)_A$ is broken to a discrete subgroup $\mathbb{Z}_q$, either $q$ should be a divisor of 192 (remember that there is a minimum of 2 zero modes in the background of a gravitational instanton) in order for $\mathbb{Z}_q$ to be free of gravitational anomaly. If $q$ is not a divisor of 192 (i.e.~$\mathbb{Z}_q$ is anomalous), a suitable TQFT must exist to saturate the anomaly. However, as we discussed in Section~\ref{subsubsec:AM_unbroken_discrete_symm}, if $q \nmid 2 \kappa_{ug}$, then no such TQFT exists on a spin manifold. 
\item mixed $\mathbb{Z}_{2}^{d\chi}$--gravity anomaly : $\kappa_{zg} = 84$ and is trivial.
\end{itemize}

\item $U(1)_A \left[SU(3)_\chi\right]^2$ anomaly: $\kappa_{u\chi^2} = 140$.
\item $\left[U(1)_A\right]^3$ anomaly: $\kappa_{u^3} = -15744$. This anomaly is milder than the mixed $U(1)_A$-gravity anomaly since $15744$ divides $192$. If $U(1)_A$ is broken to $\mathbb Z_q$, with $q$ a divisor of 192, the $\left[ \mathbb{Z}_q \right]^3$ anomaly will be automatically satisfied. 
\item $U(1)_A\left[{\rm CFU}\right]$ anomaly. For a general solution $(m,p',s)$ and setting  $n_1 = n_2 = 0$ we obtain the phase:
 \begin{eqnarray}
e^{i 2\pi\alpha\left[\frac{280 p'}{3}-15744s^2\right]}\,.
 \end{eqnarray}
 As we emphasized above, this anomaly does not depend on the color flux. For the three background fluxes we consider
\begin{equation}\label{tab:U(1)-CFU_anomaly_8_k4}
\begin{tabular}{|c|c|c|c|}
\hline
& $(4,0,0)$ & $(1,0,\frac{1}{4})$ & $(1,1,-\frac{1}{12})$ \\
 \hline
Anomaly Phase & $1$ & $e^{-i 2 \pi \alpha (984)}$ & $e^{i 2\pi\alpha (-16)}$ \\ 
\hline
\end{tabular}~\,.
\end{equation}
If the vacuum breaks $U(1)_A$ to $\mathbb Z_q$, we must have $\frac{280 p'}{3}-15744s^2=0$ mod $q$ since there is no unitary $\mathbb Z_q$-preserving TQFT on a spin manifold. This anomaly is stronger than the mixed $U(1)_A$-gravity anomaly. 
\item $\mathbb{Z}_{2}^{d\chi}\left[ SU(3)_\chi\right]^2$ anomaly: $\kappa_{z\chi^2} = 28$, and is trivial.
\item $\mathbb{Z}_{2}^{d\chi}\left[ U(1)_A\right]^2$ anomaly: $\kappa_{zu^2} = 2100$, and is trivial.
\item $\mathbb{Z}_{2}^{d\chi} \left[{\rm CFU}\right]$ anomaly:
\begin{equation}\label{tab:Zp-CFU_anomaly_8_k4}
\begin{tabular}{|c|c|c|c|}
\hline
& $(4,0,0)$ & $(1,0,\frac{1}{4})$ & $(1,1,-\frac{1}{12})$ \\
 \hline
Anomaly Phase & $1$ & $e^{i \pi}$ & $e^{i \pi}$ \\ 
\hline
\end{tabular}~\,.
\end{equation}
\end{itemize}
 $\mathbb{Z}^{d\chi}_2  \left[{\rm CFU}\right]$ anomaly is the only non-trivial discrete anomaly of the theory. Interestingly,  turning on only the  $\mathbb{Z}_2^{(1)}$  color center flux is not sufficient to see it (this is why this anomaly was missed in \cite{Bolognesi:2019fej}). We must activate more general CFU-background in order to detect it. This can also be inferred from the Dirac index of each of the three solutions: it is even for the $(4,0,0)$-background but odd for $(1,0,1/4)$ and $(1,1,-1/12)$ (see Table~\ref{tab:top_charges_Dirac_index_8_k4}).

\subsection{CFU anomalies on nonspin manifold}

We discuss the CFU-fluxes that can be activated on a nonspin manifold. We consider three solutions to the cocycle conditions eqs.~(\ref{eq:detailed cocycles psi nonspin}) and (\ref{eq:detailed cocycles chi nonspin}).
\beq
(m, p', s) = (2,0,0)~, \quad (0,0,1/2)~, \quad (1,2,1/12)~.
\eeq
In the first case we turn on only the $\mathbb{Z}_2^{(1)}$ color center, which amounts to imposing the condition $w_2 (\mathbb{M}^4) = w_2 (c)$. The second solution activates a ${\rm Spin}_c$ structure, while the third solution activates all three centers. Below, we summarize the topological charges and Dirac indices for each solution.
 \begin{equation}\label{tab:top_charges_Dirac_index_8_k4_nonspin}
\begin{tabular}{|c|c|c|c|}
\hline
& $(2,0,0)$ & $(0,0,\frac{1}{2})$ & $(1,2,\frac{1}{12})$ \\
 \hline
$Q_c$ & $\frac{7}{4}$ & $0$ & $\frac{7}{16}$ \\ 
\hline
$Q_\chi$ & $0$ & $0$ & $\frac{4}{3}$ \\ 
\hline
$Q_u$ & $0$ & $\frac{1}{8}$ &  $\frac{1}{288}$  \\ 
\hline \hline
$\mathcal{I}_\psi$ & $13$ & $360$ & $10$ \\
\hline
$\mathcal{I}_\chi$ & $21$ & $252$ & $42$ \\
\hline
\end{tabular}~\,.
\end{equation}

The UV theory has the following CFU anomalies
\begin{itemize}
\item $U(1)_A \left[{\rm CFU}\right]_{\mathbb {CP}^2}$ anomaly: 
\begin{equation}\label{tab:U(1)-CFU_anomaly_8_k4_nonspin}
\begin{tabular}{|c|c|c|c|}
\hline
& $(2,0,0)$ & $(0,0,\frac{1}{2})$ & $(1,2,\frac{1}{12})$ \\
 \hline
Anomaly Phase & $1$ & $e^{i 2\pi \alpha (-1980)}$ & $e^{i 2\pi \alpha \left( 120 \right)}$ \\ 
\hline
\end{tabular}~\,.
\end{equation}
\item $\mathbb{Z}_{2}^{d\chi}  \left[{\rm CFU} \right]_{\mathbb {CP}^2}$ anomaly:
\begin{equation}\label{tab:Zp-CFU_anomaly_8_k4_nonspin}
\begin{tabular}{|c|c|c|c|}
\hline
& $(2,0,0)$ & $(0,0,\frac{1}{2})$ & $(1,2,\frac{1}{12})$ \\
 \hline
Anomaly Phase & $e^{i \pi}$ & $1$ & $1$ \\ 
\hline
\end{tabular}~\,.
\end{equation}
Notice that turning on the $\mathbb Z_2^{(1)}$ center flux on a nonspin manifold is capable  of detecting a non-trivial $\mathbb Z_2$ phase, unlike the case on a spin manifold. In fact, we can match the $\mathbb Z_2$ phase on a nonspin manifold using a $\mathbb Z_2$-preserving TQFT \cite{Cordova:2019bsd}. 
\end{itemize}
%

\subsection{Vacuum condensates and IR phase}

We discuss anomaly matching in the IR assuming that the theory confines. 

\subsection*{Fermion bilinear condensate} 

The lowest dimensional condensate is the dimension-five fermion bilinear operator
\begin{eqnarray}
{\cal O}^{a}_1= \epsilon_{i_1ji_3i_4i_5i_6 i_7i_8}\psi^{(i_1i_2)}\chi^{a [i_3i_4i_5i_6 i_7i_8]} \left(f^c_{\mu\nu}\right)^{j}_{i_2}\sigma^{\mu\nu}~,
\label{eq:SU(8)_k4_vacuum_fermion_bilinear}
\end{eqnarray}
where ${i_{1,\cdots,8}}$ are the color indices while $a$ denotes the $SU(3)_\chi$ flavor index. The dressed gluon field $\left(f^c_{\mu\nu}\right)_{j}^{i_2}$ ensures the non-vanishing of the condensate because of statistics reasons. The expectation value of ${\cal O}^{a}_1$ breaks the symmetry $SU(3)_\chi \times U(1)_A\times \mathbb Z_2^{d\chi}$ down to $SU(2)\times U(1)\times \mathbb Z_2$.

The breaking of $SU(3)$ down to $SU(2)$ is understood by ${\cal O}^{a}_1 \propto \delta_{a1}$ using the $SU(3)$ transformation (i.e., the $SU(2)$ subgroup of $SU(3)$ can be embedded in the lower right corner of the defining $SU(3)$ matrix).
To understand the unbroken $U(1)\times \mathbb Z_2$, notice the transformations, $\chi\rightarrow -\chi$, $\psi\rightarrow \psi$ under $\mathbb Z_2^{d\chi}$ and $\chi \rightarrow e^{i 10 \pi\alpha }\chi$,  $\psi\rightarrow e^{- i18 \pi\alpha}\psi$ under $U(1)_A$ (for a continuous parameter $\alpha$) which implies ${\cal O}^{a}_1\rightarrow e^{-i 8\pi \alpha+in\pi} {\cal O}^{a}_1$ under $U(1)_A\times \mathbb Z_2^{d\chi}$ with $n=0,1$.
Additionally,  one of the $SU(3)$ Cartan generators can be taken to be $\mbox{diag}\left(e^{i 4\pi \beta}, e^{-i 2\pi \beta}, e^{-i 2\pi \beta} \right)$ for a continuous parameter $\beta$ (the other Cartan generator acts only on the $SU(2)$ subspace and is irrelevant for our discussion below).  Therefore, ${\cal O}^{a}_1$ transforms as ${\cal O}^{a}_1\rightarrow e^{i 4\pi\beta-i 8\pi \alpha+in\pi} {\cal O}^{a}_1$ under $SU(3)_\chi \times U(1)_A\times \mathbb Z_2^{d\chi}$. The direction $\beta=2\alpha\,, n=0$ remains unbroken by the condensate signaling an unbroken $U(1)$ symmetry. The direction $\alpha=\frac{1}{8}\,,\beta=0\,,n=1$ remains unbroken showing an unbroken $\mathbb Z_2$ symmetry\footnote{We thank an anonymous referee for pointing out an issue regarding the symmetry breaking pattern in our original discussion.}.

The unbroken $U(1)$ inherits the UV mixed $U(1)$-gravitational anomaly, while the unbroken $\mathbb Z_2$ inherits the UV $\mathbb Z_2^{d\chi}$[CFU] anomaly. We conclude that the bilinear condensate cannot solely match all the UV anomalies.

\subsection*{Higher-dimensional condensate}
\label{subsubsec:other_vacuum_operator}

Is it plausible to have a scenario with identically vanishing fermion bilinear operator while having a unique non-zero higher-dimensional condensate? We argue that such a possibility is unlikely. We restrict our discussion to a spin manifold where there exists no unitary TQFT that matches discrete anomalies.

A vacuum operator takes the form (up to the proper insertions of $f^c_{\mu\nu}$)
\beq
\mathcal{O} \sim \epsilon^a \psi^b \chi^c~, \label{eq:vacuum_operator_general}
\eeq
where $2b + 6c = 8 a$ to ensure gauge invariance and $b + c = 2 k$ $(k \in \mathbb{Z})$ to be a bosonic operator.
These two conditions can be combined into
\beq
k + c = 2 a~.
\label{eq:SU(8)_k4_vacuum_eq_2}
\eeq 
We organize the solutions to eq.~(\ref{eq:SU(8)_k4_vacuum_eq_2}) according to the number of fermion insertions. Notice that solution exists only for $(k,\, c) = ({\rm even},\, {\rm even})$ or $({\rm odd},\, {\rm odd})$. For $c$ even, the discrete chiral symmetry $\mathbb{Z}_{2}^{d\chi}$ is unbroken, and $\mathbb{Z}_{2}^{d\chi} \left[\mbox{CFU}\right]$ anomaly cannot be matched. Therefore, solutions with $c$ even are not acceptable. This leaves us with only those operators with odd number of both $\chi$ and $\psi$, or $c= (2\ell + 1)$ and $b = (2m+1)$ (with $\ell$, $m$ non-negative integers). 

We note that the vacuum operator in eq.~(\ref{eq:vacuum_operator_general}) has a $U(1)_A$ charge $q = -9b + 5c$, which can be shown to be a multiple of 4. This signals the breaking $U(1)_A \to \mathbb{Z}_q$. The unbroken $\mathbb{Z}_q$ needs to be anomaly-free because there exists no TQFT that can match the anomaly. We need to check that the anomalies $\left[\mathbb{Z}_q\right]^3$ (descendent from $\left[U(1)_A\right]^3$ anomaly), $\mathbb Z_q\text{-}\left[\text{grav}\right]^2$, $\mathbb Z_q\text{-} \left[\mbox{CFU}\right]$ give $0$ mod $q$. This leaves us with  $\mathbb{Z}_q$ being $\mathbb{Z}_4$ or its subgroups. The possible condensates (up to the proper insertions of $f^c_{\mu\nu}$) are
\beq
\mathcal{O} \sim \psi^1 \chi^1, \; \psi^9 \chi^{17}, \; \psi^{11} \chi^{19}, \;\; \psi^{19} \chi^{35}, \cdots
\label{eq:allowed_vacuum_SU(8)_k4}
\eeq
All these operators have charge $4$ under $U(1)_A$. The first operator is the fermion bilinear  we discussed above. We note that there exists a very large dimensional gap between the fermion bilinear  and the next higher-order operator  $\psi^9 \chi^{17}$, which has dimension $39$ (without any additional insertions of gluon fields).  Once the requirement on the $SU(3)_\chi$ anomaly matching is also included, which requires an explicit analysis of the flavor group, some of these superficially allowed operators might potentially be eliminated. Also, any extra insertions of the field strength tensor will ever increase the dimension of the operator.

\subsection*{Mixed condensates}
\label{mixed condensates}

Our analysis shows that the fermion bilinear cannot solely match all the UV anomalies. This calls for the possibility that two or more condensates are necessary to satisfy the matching conditions. Assuming that the bilinear condensate does not vanish, which breaks  $SU(3)_\chi \times U(1)_A\times \mathbb Z_2^{d\chi}$ down to $SU(2)\times U(1)\times \mathbb Z_2$,  we need at least one more condensate to match the anomalies.  One immediate candidate is ${\cal O}_2^d=\epsilon_{abc}\chi^a\chi^b\chi^c\chi^d$, which transforms as   ${\cal O}_2^d\rightarrow e^{i 40\pi \alpha+i 4\pi \beta}{\cal O}_2^d$ under $SU(3)_\chi \times U(1)_A\times \mathbb Z_2^{d\chi}$ (notice that ${\cal O}_2$ is fundamental under $SU(3)$, and therefore, it leaves an unbroken combination of $U(1)_A$ and $SU(3)$ Cartan generator). A non-vanishing ${\cal O}_2$ condensate breaks the the residual $U(1)$ and matches the mixed $U(1)$-gravitational anomaly. It also breaks the direction $\alpha=\frac{1}{8}\,,\beta=0\,, n=1$, which can be easily seen by putting $\alpha=\frac{1}{8}, \beta=0$ in the transformation factor $e^{i 40\pi \alpha+i 4\pi \beta}$ (i.e., we find ${\cal O}_2^d\rightarrow -{\cal O}_2^d$.) Therefore, a combination of the two condensates ${\cal O}_1$ (the fermion bilinear (\ref{eq:SU(8)_k4_vacuum_fermion_bilinear})) and ${\cal O}_2$ will break the residual $\mathbb Z_2$ and saturates the $\mathbb Z_2$[CFU] anomaly.

Instead of ${\cal O}_2$, one wonders whether another type of condensates, that accompanies the fermion bilinear, could match the anomalies. The answer is affirmative. Interestingly, as we shall see momentarily, the $\mathbb Z_2$[CFU] anomaly imposes constraints on the condensates more stringent than those required by the ordinary 't Hooft anomalies. Here, instead of determining all allowed operators, we simply present an example that demonstrates the extra constraint coming from the CFU anomaly. 
We consider ${\cal O}\sim \epsilon^{a_1}\psi^{a_2}\chi^{a_3}$ (with possible insertions of gluon fields). We further take ${\cal O}$ to be singlet under $SU(3)_\chi$. This makes the analysis particularly simple. Otherwise, one needs to study the possibility of having an unbroken $U(1)$ direction, a combination of $U(1)_A$ and a Cartan generator of  $SU(3)_\chi$. 
Since ${\cal O}$ is bosonic, we require $a_2+a_3=2k_1$ for some integer $k_1$. We also require ${\cal O}$ to be a color-singlet, i.e., $2a_2+6a_3=8k_2$ for some integer $k_2$.  Then, under $U(1)_A\times \mathbb Z_2^{d\chi}$ we have ${\cal O}\rightarrow e^{i2\pi \alpha (-9a_2+5a_3)+i\pi n a_3}{\cal O}$. 	Matching both $U(1)$-gravitational and $\mathbb Z_2$[CFU] mixed anomalies demands that a combination of ${\cal O}_1$ and ${\cal O}$ breaks the residual $ U(1)\times \mathbb Z_2$. We immediately find, for example, that $a_2=a_3=3$ leaves the direction $\alpha=\frac{1}{8}\,,\beta=0\,,n=1$ intact, and thus, although the condensates ${\cal O}_1$ and ${\cal O}\sim \epsilon_{abc} \psi^3\chi^a\chi^b\chi^c$ match the mixed $U(1)$-gravitational anomaly, they fail to match the new $\mathbb Z_2$[CFU] anomaly. 

\subsection*{Conformal fixed point}
\label{Conformal fixed point}

Alternatively, the theory may flow to a fixed point in the IR. We searched for a perturbative IR fixed point using the $2$-loop and $3$-loop $\beta$-function, see Appendix \ref{2loop beta function}. The 2-loop analysis does not indicate a fixed point. While the 3-loop analysis shows a fixed point with $\alpha_\text{IR} \sim 2.8$, its value is well beyond being perturbative and not reliable. Thus, one cannot trust the robustness of the fixed point. This, however, cannot exclude the possibility of a strongly-coupled CFT.

\section{Flow to an IR Conformal Field Theory}
\label{subsec:SU(8)_k2}

In this section, we analyze the chiral gauge theory $SU(8)$ with $k=2$ and discuss its possible IR phases. 
We argue that this theory is likely to flow to an (interacting) IR CFT by showing that the lowest dimensional vacuum condensate consistent with anomaly matching has dimension $\gg 4$. In addition, we show that both the $2$-loop and $3$-loop $\beta$-functions exhibit a (semi) perturbative fixed point, providing extra support to our conjecture.

\subsection{$SU(8)$ with $k=2$ ($n_\psi = 2$, $n_\chi = 6$) }

We first find the genuine discrete chiral symmetry of this theory. Since $\mbox{gcd}(N_\psi, N_\chi )=4$, one may naively conclude that the discrete symmetry  is $\mathbb Z_4$. Yet, two elements of $\mathbb Z_4$  are identified with elements in $\mathbb Z_2 \subset \mathbb Z_6$, where $\mathbb Z_6$ is the center of $SU(6)_\chi$. This leaves us with  $\mathbb Z_2^{d\chi}$ as the genuine discrete group, which we take to act solely on $\chi$. The symmetries of this theory and the charges of the fermions are summarized in the following table:
 \begin{equation}\label{tab:charges su8 with k2}
\begin{tabular}{|c|c|c|c|c|c|c|c||c|}
\hline
& $SU(2)_{\psi}$ & $SU(6)_{\chi}$ & $U(1)_{A}$ & $\mathbb Z_{2}^{d\chi}$ & $\mathbb{Z}_4^c$ & $\mathbb{Z}_2^\psi$ & $\mathbb{Z}_6^\chi$ & $\mathbb{Z}_2^{\scriptscriptstyle \rm L}$ \\
 \hline
$\psi$ & $\Box$ & $1$ & $-9$ & $0$ & $1$ & $1$ & $0$ & $1$\\ 
\hline
$\chi$ & $1$ & $\overline\Box$ & $5$ &$1$ & $-1$ & $0$  & $-1$ & $1$ \\ 
\hline
\end{tabular}~\,.
\end{equation}
In order to determine the faithful global group, we need to find the solutions to the cocycle conditions (\ref{eq:detailed cocycles psi}) and (\ref{eq:detailed cocycles chi}). There exists solutions  for every $m\in \mathbb Z_4$ with the condition $p+p'\in 2\mathbb Z$ mod $6$. This condition reduces the group elements identification (see eq. (\ref{the global structure of the global symmetry}))  from $\mathbb Z_4\times\mathbb Z_2\times \mathbb Z_6$ to $\frac{\mathbb Z_4\times\mathbb Z_2\times \mathbb Z_6}{\mathbb Z_2}\sim \mathbb Z_4\times\mathbb Z_6$. The faithful global group is, then, given by
\begin{eqnarray}
\frac{SU(2)_\psi \times SU(6)_\chi \times U(1)_A}{\mathbb Z_4\times \mathbb Z_{6}}\times \mathbb Z_2^{d\chi}\times\mathbb Z_2^{(1)}\,.
\end{eqnarray}
%

\subsection{Anomalies on spin manifold}

Let us discuss the possible background center (CFU) fluxes we can activate on  spin manifold. 
As we just mentioned, there are multiple solutions to the cocycle conditions eqs.~(\ref{eq:detailed cocycles psi}) and (\ref{eq:detailed cocycles chi}) and here we report four of them:
\beq
(m, p, p', s) = (4,0,0,0)~, \quad (3,1,3,1/4)~, \quad (2,1,3,0)~, \quad (1,1,1, 1/12)~.
\eeq
The first solution is effectively turning on only the $\mathbb{Z}_2^{(1)}$ color center, and the last one corresponds to the most refined background flux among all four.  Below, we summarize the topological charges and Dirac indices (setting $n_1=n_2=0$) for each solution.
 \begin{equation}\label{tab:top_charges_Dirac_index_8_k2_spin}
\begin{tabular}{|c|c|c|c|c|}
\hline
& $(4,0,0,0)$ & $(3,1,3,\frac{1}{4})$ & $(2,1,3,0)$ & $(1,1,1,\frac{1}{12})$ \\
 \hline
$Q_c$ & $14$ & $\frac{63}{8}$ & $\frac{7}{2}$ & $\frac{7}{8}$ \\ 
\hline
$Q_\psi$ & $0$ & $\frac{1}{2}$ & $\frac{1}{2}$ & $\frac{1}{2}$ \\ 
\hline
$Q_\chi$ & $0$ & $\frac{15}{2}$ & $\frac{15}{2}$ & $\frac{5}{6}$ \\ 
\hline
$Q_u$ & $0$ & $\frac{1}{16}$ &  $0$ & $\frac{1}{144}$  \\ 
\hline \hline
$\mathcal{I}_\psi$ & $280$ & $540$ & $88$ & $76$ \\
\hline
$\mathcal{I}_\chi$ & $504$ & $756$ & $336$ & $84$ \\
\hline
\end{tabular}~\,.
\end{equation}

The UV theory has the following `t Hooft anomalies:
\begin{itemize}
\item $\left[SU (6)_\chi\right]^3$ anomaly:  $\kappa_{\chi^3} = 28$. Unless the vacuum breaks $SU(6)_\chi$ down to an anomaly-free group, essentially there is an obstruction for the theory to confine in the IR. 
\item Mixed gravitational anomalies:
\begin{itemize}
\item Mixed $U(1)_A$-gravity anomaly : $\kappa_{ug} = 192$. 
\item Mixed $\mathbb{Z}_{2}^{d\chi}$--gravity anomaly : $\kappa_{zg} = 168$ and is trivial.
\end{itemize}

\item $U(1)_A\left[SU(6)_\chi\right]^2$ anomaly: $\kappa_{u\chi^2} = 140$.
\item $U(1)_A \left[SU(2)_\psi\right]^2$ anomaly: $\kappa_{u\psi^2} = -324$.
\item $\left[U(1)_A\right]^3$ anomaly: $\kappa_{u^3} = -31488$
\item $U(1)_A \left[{\rm CFU}\right]$ anomaly (setting $n_1 = n_2 = 0$):
\begin{equation}\label{tab:U(1)-CFU_anomaly_8_k2_spin}
\begin{tabular}{|c|c|c|c|c|}
\hline
& $(4,0,0,0)$ & $(3,1,3,\frac{1}{4})$ & $(2,1,3,0)$ & $(1,1,1,\frac{1}{12})$ \\
 \hline
Anomaly Phase & $1$ & $e^{i 2\pi \alpha (-1080)}$ & $e^{i 2\pi \alpha (888)}$ & $e^{i 2\pi \alpha (-264)}$ \\ 
\hline
\end{tabular}~\,.
\end{equation}
\item $\mathbb{Z}_{2}^{d\chi}  \left[SU(6)_\chi\right]^2$ anomaly: $\kappa_{z\chi^2} = 28$, and is trivial.
\item $\mathbb{Z}_{2}^{d\chi}  \left[U(1)_A\right]^2$ anomaly: $\kappa_{zu^2} = 4200$, so is trivial.
\item $\mathbb{Z}_{2}^{d\chi} \left[ {\rm CFU}\right]$ anomaly:
\begin{equation}\label{tab:Zp-CFU_anomaly_8_k2_spin}
\begin{tabular}{|c|c|c|c|c|}
\hline
& $(4,0,0,0)$ & $(3,1,3,\frac{1}{4})$ & $(2,1,3,0)$ & $(1,1,1,\frac{1}{12})$ \\
 \hline
Anomaly Phase & $1$ & $1$ & $1$ & $1$ \\ 
\hline
\end{tabular}~\,.
\end{equation}
We see that there is no non-trivial  $\mathbb{Z}_{2}^{d\chi} \left[ {\rm CFU}\right]$ anomaly on spin manifold. In order to find a non-trivial discrete anomaly, we need to activate even more refined background fluxes, and the only remaining option is to place the theory on a nonspin manifold.
\end{itemize}

All our comments from the previous section still apply when we consider the breaking of $U(1)_A$ to a discrete group in the gravitational or CFU backgrounds.

\subsection{CFU anomalies on nonspin manifold}

We discuss possible background center (CFU) fluxes we can activate on nonspin manifold. 
There are multiple solutions to the cocycle conditions in eqs.~(\ref{eq:detailed cocycles psi nonspin}) and (\ref{eq:detailed cocycles chi nonspin}) and here we report four of them:
\beq
(m, p, p', s) = (2,0,0,0)~, \quad (0,0,0,1/2)~, \quad (0,1,3,0)~, \quad (1,1,1, -5/12)~.
\eeq
The first solution is effectively turning on only $\mathbb{Z}_2^{(1)}$ color center with the condition $w_2 (\mathbb{M}^4) = w_2 (c)$, and the second configuration is nothing but ${\rm Spin}_c$ manifold. The third configuration may be viewed as a generalization of non-spin manifold case occurred in $SU(2)$ adjoint QCD. Here, $w_2 (\mathbb{M}^4)$ of the tangent bundle is balanced by a combination of two center fluxes of the flavor groups $SU(2)_\psi \times SU(6)_\chi$.  Below, we summarize the topological charges and Dirac indices  for each solution.
 \begin{equation}\label{tab:top_charges_Dirac_index_8_k2_nonspin}
\begin{tabular}{|c|c|c|c|c|}
\hline
& $(2,0,0,0)$ & $(0,0,0,1/2)$ & $(0,1,3,0)$ & $(1,1,1, -5/12)$ \\
 \hline
$Q_c$ & $\frac{7}{4}$ & $0$ & $0$ & $\frac{7}{16}$ \\ 
\hline
$Q_\psi$ & $0$ & $0$ & $\frac{1}{4}$ & $\frac{1}{4}$ \\ 
\hline
$Q_\chi$ & $0$ & $0$ & $\frac{15}{4}$ & $\frac{5}{12}$ \\ 
\hline
$Q_u$ & $0$ & $\frac{1}{8}$ &  $0$ & $\frac{25}{288}$  \\ 
\hline
$Q_g$ & $-\frac{1}{8}$ & $-\frac{1}{8}$ &  $-\frac{1}{8}$ & $-\frac{1}{8}$  \\ 
\hline \hline
$\mathcal{I}_\psi$ & $26$ & $720$ & $0$ & $515$ \\
\hline
$\mathcal{I}_\chi$ & $42$ & $504$ & $84$ & $371$ \\
\hline
\end{tabular}~\,.
\end{equation}

The UV theory has the following CFU anomalies.
\begin{itemize}
\item $U(1)_A \left[{\rm CFU}\right]_{\mathbb{ CP}^2}$ anomaly (with $n_1 = n_2 = 0$):
\begin{equation}\label{tab:U(1)-CFU_anomaly_8_k2_nonspin}
\begin{tabular}{|c|c|c|c|c|}
\hline
& $(2,0,0,0)$ & $(0,0,0,\frac{1}{2})$ & $(0,1,3,0)$ & $(1,1,1, -\frac{5}{12})$ \\
 \hline
Anomaly Phase & $e^{i 2\pi \alpha (-24)}$ & $e^{i 2\pi \alpha (-3960)}$ & $e^{i 2\pi \alpha (420)}$ & $e^{i 2\pi \alpha (-2780)}$ \\ 
\hline
\end{tabular}~\,.
\end{equation}
\item $\mathbb{Z}_{2}^{d\chi}  \left[{\rm CFU}\right]_{\mathbb{CP}^2}$ anomaly:
\begin{equation}\label{tab:Zp-CFU_anomaly_8_k2_nonspin}
\begin{tabular}{|c|c|c|c|c|}
\hline
& $(2,0,0,0)$ & $(0,0,0,\frac{1}{2})$ & $(0,1,3,0)$ & $(1,1,1, -\frac{5}{12})$\\
 \hline
Anomaly Phase & $1$ & $1$ & $1$ & $e^{i\pi}$ \\ 
\hline
\end{tabular}~\,.
\end{equation}
\end{itemize}
There is now a non-trivial $\mathbb{Z}_{2}^{d\chi} \left[{\rm CFU}\right]_{\mathbb {CP}^2}$ anomaly in the background of the CFU configuration $(m,p,p',s) = (1,1,1, -\frac{5}{12})$.

\subsection{Vacuum condensates and IR phase}

We finally discuss anomaly matching in the IR assuming the theory confines. 

\subsection*{Fermion bilnear condensate} 
The lowest dimensional color-singlet condensate is 
\begin{eqnarray}
{\cal O}^{A}_a= \epsilon_{i_1ji_3i_4i_5i_6 i_7i_8}\psi^{A(i_1i_2)}\chi^{ [i_3i_4i_5i_6 i_7i_8]}_a \left(f^c_{\mu\nu}\right)^{j}_{i_2}\sigma^{\mu\nu}\,,
\end{eqnarray}
where  $a=1,2,...,6$ are $SU(6)_\chi$ and $A=1,2$ are $SU(2)_\psi$ flavor indices. This condensate breaks $\mathbb Z_2\rightarrow 1$ (remember that $\mathbb Z_2$ is chosen to act solely on $\chi$), which is sufficient to match the $\mathbb Z_2^{d\chi}\left[\mbox{CFU}\right]_{\mathbb {CP}^2}$ anomaly. The condensate, however, exhibits a sever problem that excludes it as a possible IR scenario. Since ${\cal O}^{A}_a$ transforms in the defining representation of $SU(6)_\chi \times SU(2)_\psi$, it breaks it down to either $H\supset SU(5)\times U(1)$ or $H\supset SU(4)\times SU(2)$ (in the second case, $SU(2)$ is a vector-like group; see  \cite{Li:1973mq}). Both choices of $H$ are not anomaly free, as can be easily seen. A traditional way out of this problem would be the formation of composite massless fermions in the IR that are charged under $H$. This solution is excluded, thanks to the fact that the theory at hand does not admit fermions in its spectrum.

\subsection*{Higher-dimensional condensates}

We can repeat the analysis presented in Section~\ref{subsubsec:other_vacuum_operator}, with a sole difference that the vacuum operator needs not break $\mathbb{Z}_2^{d\chi}$. The reason is the following. First, we have shown that there is no non-trivial mixed $\mathbb Z_2^{d\chi}\left[\mbox{CFU}\right]$ anomaly on spin manifold, removing the requirement of anomaly matching. Then, on nonspin manifold, while there is a $\mathbb Z_2^{d\chi}\left[\mbox{CFU}\right]_{\mathbb{ CP}^2}$ anomaly, it needs not be matched by breaking $\mathbb{Z}_2^{d\chi}$ since a $\mathbb Z_2$-preserving TQFT exists on $\mathbb {CP}^2$. 

Relaxing the requirement of $\mathbb{Z}_2^{d\chi}$ breaking, we again find that the general vacuum operator has $U(1)_A$ charge of multiples of 4. In addition, the maximal anomaly-free subgroup of $U(1)_A$ is shown to be $\mathbb{Z}_4$ on both spin and non-spin manifolds. The allowed condensates are given by
\beq
\mathcal{O} \sim \psi^1 \chi^1, \; \psi^4 \chi^8, \; \psi^6 \chi^{10}, \; \psi^9 \chi^{17}, \; \psi^{11} \chi^{19}, \; \psi^{14} \chi^{26}, \cdots.
\eeq
As anticipated, this time we have solutions with even number of $\chi$'s in addition to the ones listed in eq.~(\ref{eq:allowed_vacuum_SU(8)_k4}). We first recall that the leading operator $\mathcal{O} \sim \psi \chi$ is what we studied above and we showed that it is ruled out by anomaly matching of continuous symmetries. This means that the lowest dimensional vacuum condensate consistent with $U(1)_A$ anomaly matching is a 12-fermion operator (possibly with additional $f^c_{\mu\nu}$ insertions). We also need to check if this operator can saturate the continuous anomaly of $SU(6)_\chi$. If this is the case, then the operator $ \psi^4 \chi^8$ might have a nonvanishing expectation value and be accompanied by a $\mathbb Z_2$ TQFT.  However, given that this operator has a scaling dimension of at least 18 (zero insertions of $f^c_{\mu\nu}$), we conjecture that there is some  chance for this theory to avoid confinement and flow to a CFT. Rather surprisingly, this conclusion is in fact consistent with the $2$-loop and $3$-loop $\beta$-function calculations. The $2$-loop calculations yield a value of the fixed-point coupling constant $\alpha_{\scriptscriptstyle \rm IR}= \frac{64 \pi}{659}\approx 0.305$, while the $3$-loop calculations yield $\alpha_{\scriptscriptstyle \rm IR}=0.181$,  see  Appendix \ref{beta function}. 

However, this is a speculative conclusion rather than one based on firm calculations; after all, we are discussing a strongly-coupled theory. Some caveats against a CFT are as follows: 

\begin{enumerate}

\item It is well known that certain supersymmetric field theories with  a number of fundamental chiral fields  and  chiral adjoint field $X$ with superpotential $W=\tr X^{k+1}$, and $k$ very large, can exhibit symmetry breaking and non-trivial IR dynamics, see \cite{Kutasov:1995np}. Such operators are said to be dangerously irrelevant since they receive non-vanishing vacuum expectation values  irrespective of their large scaling dimensions. However, this class of theories fundamentally differs from the theories we discuss here, thanks to the extra scalars and, above all, the holomorphy of supersymmetry.

\item One may argue that a set of operators $\sim\psi\chi f^c,\psi\chi \left(f^{c}\right)^2,\psi\chi \left(f^{c}\right)^3$, etc. may break $SU(6)_\chi\times SU(2)_\psi\times U(1)_A$ to an anomaly-free subgroup. We cannot exclude this possibility. However, given the large scaling dimension of these operators and the fact that all of them are aligned in flavor space, we find that such scenario is less likely.

\item As we pointed out in the previous section, multiple condensates that are not aligned in flavor space may break the symmetries to an anomaly-free subgroup. The analysis of the multi-condensates, however, is more involved and beyond the scope of this paper. For example, the the condensate $\psi\chi$ will break the global symmetry $SU(6)_\chi\times SU(2)_\psi\times U(1)_A$ down to either $ H\supset SU(5)\times U(1)$ or $ H \supset SU(4)\times SU(2)$. Both of these symmetry breaking patterns are anomalous. However, the condensate $\chi\chi\chi\chi$ might further break one of them to an anomaly-free subgroup. The operator $\chi\chi\chi\chi$ transforms in the 4-index symmetric representation and it is not immediately clear how it may break the original group $SU(6)$ or one of the unbroken subgroups $SU(5)$ or $SU(4)$ to anomaly-free parts.  A future work should address such interesting possibility.

\end{enumerate}

{\flushleft \bf Acknowledgments.} We would like to thank T.~Daniel Brennan, Clay Cordova, Erich Poppitz, Mithat \"{U}nsal, and Liantao Wang for many enlightening discussions and especially T.~Daniel Brennan and Erich Poppitz for comments on the manuscript. 
The work of M.A. has been supported in part by NSF grant PHY-2013827 and in part by STFC through grant ST/T000708/1.  S.H. is supported by a DOE grant DE-SC-0013642 and a DOE grant DE-AC02-06CH11357. MS was supported by the Samsung Science and Technology Foundation under Project Number SSTF-BA1602-04.

\appendix

\section{The Independent Discrete Symmetry}
\label{The discrete symmetry}

In this appendix we show that the global structure of $U(1)_A \times \mathbb Z_{N_{\psi}+N_\chi}$ is given by
\beq
\frac{U(1)_A \times \mathbb Z_{N_{\psi}+N_\chi}}{Z_{(N_{\psi}+N_\chi)/r}}~, \quad r = \mbox{gcd}(N_\psi,N_\chi)
\label{eq:global_structure_U(1)_Zr}~,
\eeq
hence showing that there are $r$ independent discrete symmetry generators. 

We recall that the symmetry $\mathbb Z_{N_\psi+N_{\chi}}$ acts simultaneously on both $\psi$ and $\chi$ with common charge 1. Then, we notice that the action of $\mathbb Z_{N_{\psi}+N_\chi}$ on $\psi$  can be undone by appropriately choosing the parameter of the  $U(1)_A$. Thus, under the action of $U(1)_A\times \mathbb Z_{N_\psi+N_{\chi}}$ we have
\begin{eqnarray}
\psi=e^{i2 \pi\alpha a q_{\chi}}e^{i\frac{2\pi \ell}{N_\psi+N_{\chi}}}\psi~,
\end{eqnarray}
for $\ell=1,2,..., N_\psi+N_{\chi}$. Here, $q_\psi = - N_\chi /r$ and we have included the charge scaling factor $a$, which we use to show that the global structure (and the number of independent generator) is insensitive to the charge assignment. This yields the solution
\begin{eqnarray}
\alpha = - \frac{\ell}{a q_{\chi}(N_\psi+N_{\chi})} + \frac{k_1}{a q_{\chi}}~,
\end{eqnarray}
and $k_1 \in \mathbb Z$.
Next, we use the parameter $\alpha$ to find  the action of $U(1)_{A}\times \mathbb Z_{N_\psi+N_{\chi}}$ on $\chi$:
\begin{eqnarray}
\chi \rightarrow e^{i 2\pi a q_\chi \alpha} e^{i \frac{2\pi \ell}{N_\psi + N_\chi}}  = e^{i2\pi \frac{(\ell-k_1 N_{\psi})}{N_{\chi}}}\chi~,
\end{eqnarray}
where $q_\chi = N_\psi /r$. We note that $a$-dependence is completely dropped out, and the results obtained below are therefore insensitive to the charge assignment.
Now, any integer $\ell$ can be written as $\ell=m_1+m_2 r$ for $m_2\in \mathbb Z$ and $m_1=1,2, \cdots , r-1$. Thus, the transformation on $\chi$ becomes
\begin{eqnarray}
\chi \rightarrow e^{i2\pi \frac{(m_1+m_2 r - k_1 N_{\psi})}{N_{\chi}}}\chi~.
\end{eqnarray}
Next we use Bezout's identity, which states that there exists an integer $k_2$ such that 
\begin{eqnarray}
m_2\, \mbox{gcd}(N_\psi,N_{\chi})=k_1 N_{\psi}+k_2N_{\chi}~,
\end{eqnarray}
which immediately yields
\begin{eqnarray}
\chi \rightarrow e^{i2\pi \frac{m_1}{N_{\chi}}}\chi~,
\label{eq:indep_Zr_transformation_chi}
\end{eqnarray}
with generators $m_1=1,2 \cdots ,r-1$. Thus, there are only $r = \mbox{gcd}(N_\psi,N_{\chi})$ independent generators, including the identity generator, that faithfully act on $\chi$. In other words, there are $\mathbb Z_r$ orbits inside  $\frac{U(1)_A \times \mathbb Z_{N_{\psi}+N_\chi}}{Z_{(N_{\psi}+N_\chi)/r}}$. The action of $\mathbb Z_r$ on $\chi$ is then represented as
\begin{eqnarray}
\chi \rightarrow e^{i2\pi \frac{m}{r}}\chi \,,\quad m=0,1,...,r-1~.
\end{eqnarray}
%

\section{The Descend Procedure, Reducible Anomalies, and Counter Terms}
\label{The descend procedure and counter terms}

In this appendix we describe the descend procedure that produces the anomalies as we descend from $6$ to $4$ dimensions as well as the possibility of adding counter terms whenever we have a reducible anomaly . As an example, consider a theory with $n_{\chi}$ flavors of fermions $\chi$ and a single flavor $n_{\psi}=1$  of fermion $\psi$. The choice  $n_{\psi}=1$ is made in order to reduce clutter\footnote{This prototype example is the case $SU(8)$ with $k=4$.}. Otherwise, generalizing the treatment to a general number of flavors is a straightforward task. The theory admits the following classical symmetry 
\begin{eqnarray}
G= SU(N)\times SU(n_\chi)\times U(1)_1\times U(1)_2~,
\end{eqnarray}
and we take the actions of $U(1)_1$ and $U(1)_2$ on $\psi$ and $\chi$ to be
\begin{eqnarray}
\nonumber
U(1)_1&:& \quad \psi \rightarrow e^{i \alpha_1}\psi\,,\quad \chi \rightarrow e^{i\beta_1}\chi~,
\\[3pt]
U(1)_2&:& \quad \psi \rightarrow e^{i \alpha_2}\psi\,,\quad \chi \rightarrow e^{i\beta_2}\chi~.
\end{eqnarray}
First, we would like to determine the nonanomalous part of $G$. To this end, we turn on background flux fields $f^{c}_2$, $F^{f}_2$,  $F^{u_1}_2$, and  $F^{u_2}_2$ of $SU(N)$, $SU(n_f=n_\chi)$, $U(1)_1$, and $U(1)_2$, respectively. Then, we  write down the anomaly polynomial in $6$ dimensions as:
\begin{eqnarray}
\Omega_{6}=\frac{2\pi}{3!}\left(\frac{-iF^t_2}{2\pi}\right)^3~,
\label{full anomaly polynomial}
\end{eqnarray}
where $F^t_2$ is a superposition of all  background fields:
\begin{eqnarray}
 F^t_2=f_2^c\otimes \mathbf{1}_f+ \mathbf{1}_c\otimes F_2^f+ \mathbf{1}_c\otimes \mathbf{1}_f \otimes \left(F^{u_1}_2+F^{u_2}_2\right)~.
 \label{all fields}
\end{eqnarray}
Notice that we kept traces over non-abelian fields implicit. 
Substituting (\ref{all fields}) into (\ref{full anomaly polynomial}), keeping all possible terms that account for both pure and mixed anomalies, we find:
\begin{eqnarray}
\nonumber
\Omega_{6}&=&\frac{1}{(2\pi)^2}\left[\frac{1}{3!}\kappa_{c^3}f^{c}_2\wedge f^{c}_2\wedge f^{c}_2+\frac{1}{2!}\kappa_{u_1c^2}  F^{u_1}_2\wedge f^{c}_2  \wedge f^{c}_2+ \frac{1}{2!}\kappa_{u_2c^2}  F^{u_2}_2\wedge f^{c}_2  \wedge f^{c}_2\right.
\\[3pt]
\nonumber
&&+\frac{1}{2!}\kappa_{u_1f^2}  F^{u_1}_2\wedge F^{f}_2  \wedge F^{f}_2+\frac{1}{2!}\kappa_{u_2f^2}  F^{u_2}_2\wedge F^{f}_2  \wedge F^{f}_2+\frac{1}{2!}\kappa_{u_1u_2^2}  F^{u_1}_2\wedge F^{u_2}_2  \wedge F^{u_2}_2
\\[3pt]
\nonumber
&&+\frac{1}{2!}\kappa_{u_2u_1^2}  F^{u_2}_2\wedge F^{u_1}_2  \wedge F^{u_1}_2+\frac{1}{3!}\kappa_{u_1^3}  F^{u_1}_2\wedge F^{u_1}_2  \wedge F^{u_1}_2
\\[3pt]
&&\left.+\frac{1}{3!}\kappa_{u_2^3}  F^{u_2}_2\wedge F^{u_2}_2  \wedge F^{u_2}_2+\frac{1}{3!}\kappa_{f^3}F^{f}_2\wedge F^{f}_2\wedge F^{f}_2\right]\,.
\label{very general anomaly}
\end{eqnarray}
Notice that terms like $f^{c}_2\wedge F^{u_1}_2\wedge F^{u_1}_2$ are zero by construction since $\mbox{tr}[t^a]=0$ for the non-abelian Lie-algebra generators $t^a$. 

Next, recall that an anomaly $\Omega_{d+2}^{\scriptsize\mbox{red}}$ is said to be reducible if we can write $\Omega_{d+2}^{\scriptsize\mbox{red}}$  as
\begin{eqnarray}
\Omega_{d+2}^{\scriptsize\mbox{red}}={\cal J}_{p}\wedge {\cal K}_{d+2-p}~,
\end{eqnarray}
and ${\cal J}_{p}$, ${\cal K}_{d+2-p}$ satisfy
\begin{eqnarray}
d{\cal J}_{p}=0\,,\quad d{\cal K}_{d+2-p}=0~.
\end{eqnarray}
Then, we  write $\Omega_{d+2}^{\scriptsize\mbox{red}}$, at least locally, as 
\begin{eqnarray}
\Omega_{d+2}^{\scriptsize\mbox{red}}=d \omega_{d+1}^{\scriptsize\mbox{red}}~,
\end{eqnarray}
where 
\begin{eqnarray}
\omega_{d+1}^{\scriptsize\mbox{red}}={\cal J}_{p-1}\wedge {\cal K}_{d+2-p}+sd\left({\cal J}_{p-1}\wedge {\cal K}_{d+1-p}\right)~,\quad s\in \mathbb R
\label{counter term appearance}
\end{eqnarray}
and ${\cal J}_{p}$, ${\cal K}_{d+2-p}$ satisfy ${\cal J}_{p}=d{\cal J}_{p-1}$, ${\cal K}_{d+2-p}=d{\cal K}_{d+1-p}$, respectively. The last term in (\ref{counter term appearance}) is a counter term, which can always be added as we descend from $6$ to $5$ dimensions. 

All mixed anomalies in (\ref{very general anomaly}) are reducible. Consider, as an example, the term 
\begin{eqnarray}
\frac{1}{2!}\kappa_{u_1c^2}  F^{u_1}_2\wedge f^{c}_2  \wedge f^{c}_2~.
\end{eqnarray}
Then, we can identify ${\cal J}_{2}$ and ${\cal K}_{4}$ as
\begin{eqnarray}
{\cal J}_{2}=F^{u_1}_2\,,\quad {\cal K}_{4}=f^{c}_2\wedge f^{c}_2
\end{eqnarray}
since 
\begin{eqnarray}
\nonumber
 &&d F^{u_1}_2=dd A^{u_1}_1=0~,
 \\[3pt]
 \nonumber
&&d\left[ f^{c}_2\wedge f^{c}_2\right]=d d\mbox{CS}(a^{c}_1)=dd\left[a^{c}_1\wedge d a^{c}_1+\frac{2}{3}a^{c}_1\wedge a^{c}_1\wedge a^{c}_1\right]=0~.
 \end{eqnarray}
Thus, we immediately conclude
\begin{eqnarray}
{\cal J}_{1}=A^{u_1}_1~, \quad {\cal K}^{}_3=\mbox{CS}(a^{c}_1)\equiv a^{c}_1\wedge da^{c}_1+\frac{2}{3}a^{c}_1\wedge a^{c}_1\wedge a^{c}_1~.
\end{eqnarray}
Then, upon descending from $6$ to $5$ dimensions we obtain:
\begin{eqnarray}
\begin{split}
\frac{1}{2!}\kappa_{u_1c^2}  F^{u_1}_2\wedge f^{c}_2  \wedge f^{c}_2\longrightarrow& \frac{1}{2!}\kappa_{u_1c^2} A^{u_1}_1\wedge f^{c}_2\wedge f^{c}_2
\\[3pt]
&+s\, d\Big [ A^{u_1}\wedge \mbox{CS}(a^{c}_1) \Big ]~.
\end{split}
\end{eqnarray}

We can apply the same analysis to all the mixed terms in  (\ref{very general anomaly}). The final 5$d$ anomaly polynomial is
\begin{eqnarray}
\begin{split}
(2\pi)^2 \omega_{5} &= \frac{1}{2!}\kappa_{u_1 c^2} A^{u_1}_1\wedge f^{c}_2\wedge f^{c}_2
+s_1 d \Big [ A^{u_1}_1\wedge \mbox{CS}(a^{c}_1) \Big ]+ \frac{1}{2!}\kappa_{u_2 c^2} A^{u_2}_1\wedge f^{c}_2\wedge f^{c}_2
\\[3pt]
&+s_2 d \Big [ A^{u_2}_1\wedge \mbox{CS}(a^{c}_1)\Big ]+\frac{1}{2!}\kappa_{u_1f^2} A^{u_1}_1\wedge F^{f}_2\wedge F^{f}_2+r_1 d \Big  [A^{u_1}_1\wedge \mbox{CS}(A^{f}_1)\Big ]
\\[3pt]
&+ \frac{1}{2!}\kappa_{u_2f^2} A^{u_2}_1\wedge F^{f}_2\wedge F^{f}_2
+ r_2 d \Big [A^{u_2}_1\wedge \mbox{CS}(A^{f}_1)\Big ]+\frac{1}{2!}\kappa_{u_1u_2^2} A^{u_1}_1\wedge F^{u_2}_2\wedge F^{u_2}_2\\
&+t_1 d\Big [A^{u_1}_1\wedge A^{u_2}_1\wedge F^{u_2}_2 \Big ]+\frac{1}{2!}\kappa_{u_2u_1^2} A^{u_2}_1\wedge F^{u_1}_2\wedge F^{u_1}_2+t_2 d\Big [A^{u_2 }_1\wedge A^{u_1 }_1\wedge F^{u_1}_2 \Big ]
\\[3pt]
&+\frac{1}{3!}\kappa_{c^3}Q_5^c+\frac{1}{3!}\kappa_{f^3}Q_5^f+\frac{1}{3!}\kappa_{u_1^3} Q_5^{u_1}
+\frac{1}{3!}\kappa_{u_2^3}  Q_5^{u_2}~,
\end{split}
\end{eqnarray}
where $Q_5=A_1\wedge F_2\wedge F_2-\frac{A_1^3\wedge F_2}{2}+\frac{A_1^5}{10}$ for the non-abelian fields and $Q_5=A_1\wedge F_2^2$ for the abelian fields. 
We perform the gauge transformations $A^{u_1}_1\rightarrow A^{u_1 }_1+d\lambda^{u_1}_0$, $A^{u_2 }_1\rightarrow A^{u_2 }_1+d\lambda^{u_2}_0$, $\mbox{CS}(a^{c}_1)\rightarrow \mbox{CS}(a^{c}_1)+d\left(\lambda^{c}_0 da^{c}_1\right)$,  $\mbox{CS}(A^{f}_1)\rightarrow \mbox{CS}(A^{f}_1)+d (\lambda^{f}_0 dA^{f}_1)$, and $Q_5\rightarrow Q_5+dQ_4$, where $Q_4=\lambda_0 d(A_1\wedge dA_1+A_1^3/2)$ for the non-abelian fields and $Q_4=\lambda_0 d(A_1\wedge dA_1)$ for the abelian fields. We also assume that the gauge parameters $\lambda^{u_1}_0$, $\lambda^{u_2}_0$, $\lambda^{c}_0$, and $\lambda^{f}_0$  have local supports on a closed $4$-dimensional manifold ${\mathbb M}^4$. This procedure produces all the $4$-dimensional anomalies including the counter terms\footnote{Notice that this procedure gives the consistent anomaly.}:
\begin{eqnarray}
\begin{split}
{\cal A}^c=&\frac{i}{4\pi^2}\int_{{\mathbb M}^4} \lambda^{c}_0\left[\frac{\kappa_{c^3}}{3!} d\left(a_1^c\wedge da_1^c+\frac{(a_1^c)^3}{2}\right) +s_1 F^{u_1}_2\wedge d a^{c}_1+s_2 F^{u_2}_2\wedge da^{c}_1\right]~,
\\[3pt]
{\cal A}^f =&\frac{i}{4\pi^2}\int_{{\mathbb M}^4} \lambda^{f}_0\left[\frac{\kappa_{f^3}}{3!}  d\left(A_1^f\wedge dA_1^f+\frac{(A_1^f)^3}{2}\right) +r_1 F^{u_1 }_2\wedge dA^{f}_1+r_2 F^{u_2 }_2\wedge dA^{f}_1\right]~,
\\[3pt]
{\cal A}^{u_1} =&\frac{i}{4\pi^2}\int_{{\mathbb  M}^4} \lambda^{u_1}_0\left[\frac{\kappa_{u_1^3}}{3!}  F^{u_1}_2\wedge F^{u_1}_2+\left(\frac{\kappa_{u_1c^2}}{2!}-s_1\right)  f^{c}_2\wedge f^{c}_2\right.
\\[3pt]
&\left. + \left(\frac{\kappa_{u_1f^2}}{2!}-r_1\right)  F^{f}_2\wedge F^{f}_2+ \left(\frac{\kappa_{u_1u_2^2}}{2!}-t_1\right)  F^{u_2}_2\wedge F^{u_2}_2 +t_2  F^{u_1}_2\wedge F^{u_2}_2 \right]~,
\\[3pt]
{\cal A}^{u_2}=&\frac{i}{4\pi^2}\int_{{\mathbb M}^4} \lambda^{u_2}_0\left[\frac{\kappa_{u_2^3}}{3!}  F^{u_2}_2\wedge F^{u_2}_2+\left(\frac{\kappa_{u_2c^2}}{2!}-s_2\right)  f^{c}_2\wedge f^{c}_2\right.
\\[3pt]
&\left. + \left(\frac{\kappa_{u_2f^2}}{2!}-r_2\right)  F^{f}_2\wedge F^{f}_2+ \left(\frac{\kappa_{u_2u_1^2}}{2!}-t_2\right)  F^{u_1}_2\wedge F^{u_1}_2 +t_1  F^{u_1}_2\wedge F^{u_2}_2 \right]~.
\end{split}
\end{eqnarray}
Since we are gauging $SU(N)$, the anomaly ${\cal A}^c$ has to vanish identically. This is achieved by having an appropriate number of Weyl fermions $\psi$ and $\chi$ and, at the same time, by setting the counter terms $s_1$ and $s_2$ to zero. The rest of the counter terms $r_1, r_2, t_1,t_2$ can be chosen arbitrarily. This has the effect of shuffling the mixed anomalies around. A canonical choice is to set $r_1=r_2= t_1=t_2=0$. This leaves us with the anomalies:
\begin{eqnarray}
\nonumber
{\cal A}^f&=&\frac{i}{4\pi^2}\int_{{\mathbb M}^4} \lambda^{f}_0\left[\frac{\kappa_{f^3}}{3!}   d\left(A_1^f\wedge dA_1^f+\frac{(A_1^f)^3}{2}\right) \right]~,
\\[3pt]
\nonumber
{\cal A}^{u_1}&=&\frac{i}{4\pi^2}\int_{{\mathbb M}^4} \lambda^{u_1}_0\left[\frac{\kappa_{u_1^3}}{3!}  F^{u_1}_2\wedge F^{u_1}_2+\frac{\kappa_{u_1c^2}}{2!}  f^{c}_2\wedge f^{c}_2 + \frac{\kappa_{u_1f^2}}{2!}  F^{f}_2\wedge F^{f}_2+\frac{\kappa_{u_1u_2^2}}{2!}  F^{u_2}_2\wedge F^{u_2}_2 \right]~,
\\[3pt]
\nonumber
{\cal A}^{u_2}&=&\frac{i}{4\pi^2}\int_{{\mathbb M}^4} \lambda^{u_2}_0\left[\frac{\kappa_{u_2^3}}{3!}  F^{u_2}_2\wedge F^{u_2}_2+\frac{\kappa_{u_2c^2}}{2!}  f^{c}_2\wedge f^{c}_2+ \frac{\kappa_{u_2f^2}}{2!}  F^{f}_2\wedge F^{f}_2+ \frac{\kappa_{u_2u_1^2}}{2!}  F^{u_1}_2\wedge F^{u_1}_2 \right]~,
\end{eqnarray}
where the anomaly coefficients are given by
\begin{eqnarray}
\begin{split}
&\kappa_{f^3}=\mbox{dim}_\chi\,,\quad \kappa_{u_1^3}=\alpha_1\mbox{dim}_\psi+ n_\chi\beta_1\mbox{dim}_\chi\,,\quad \kappa_{u_2^3}=\alpha_2\mbox{dim}_\psi+ n_\chi\beta_2\mbox{dim}_\chi~,
\\[3pt]
&\kappa_{u_1c^2}=\alpha_1 T_{\psi}+n_\chi \beta_1 T_{\chi}\,,\quad \kappa_{u_2c^2}=\alpha_2 T_{\psi}+n_\chi \beta_2 T_{\chi}~,
\\[3pt]
&\kappa_{u_1f^2}=\beta_1 \mbox{dim}_\chi\,,\quad \kappa_{u_2f^2}=\alpha_2 \mbox{dim}_\chi~.
\end{split}
\end{eqnarray}
In particular, we can find a combination of the gauge parameters $\alpha_1$, $\alpha_2$, $\beta_1$, and $\beta_2$ that leaves  $U(1)_1$ anomaly free in the color instanton background. We call this symmetry the axial $U(1)_A$:
\begin{eqnarray}
U(1)_A: \psi\rightarrow e^{-i2\pi \frac{n_\chi T_\chi}{r}}\psi\,, \quad \chi \rightarrow  e^{i2\pi \frac{T_\psi}{r}}\chi\,,
\end{eqnarray}
and $r=\mbox{gcd}(n_\chi T_\chi,\, T_\psi)$. We can also find special values of $\alpha_2$ and $\beta_2$ that leaves a discrete subgroup of $U(1)_2$ invariant in the color background. This is $\mathbb Z_{n_\chi T_\chi+T_\psi}$ that acts on the fermions as
\begin{eqnarray}
\mathbb Z_{n_\chi T_\chi+T_\psi}: \psi\rightarrow e^{i 2\pi \frac{\ell}{n_\chi T_\chi+T_\psi}} \psi\,,\quad \chi \rightarrow e^{i 2\pi \frac{\ell}{n_\chi T_\chi+T_\psi}}\chi~,
\end{eqnarray}
where $\ell=0,1,...,n_\chi T_\chi+T_\psi-1$.
However, as we discussed in the bulk of the paper and in Appendix \ref{The discrete symmetry}, only a discrete subgroup $\mathbb Z_{\scriptsize\mbox{gcd}(n_\chi T_\chi, T_\psi)}\subset \mathbb Z_{n_\chi T_\chi+T_\psi}$ acts faithfully on fermions.

Now, we may restore the general number of flavors $n_\psi$. We recall  that the $U(1)_A$ charges are given by $q_\psi=-\frac{n_\chi T_\chi}{r}$ and $q_\chi= \frac{n_\psi T_\psi}{r}$, where $r=\mbox{gcd}(n_\chi T_\chi,n_\psi T_\psi)$. Then,  the following is the  set of complete traditional (perturbative) 't Hooft anomalies: 
 \begin{eqnarray}\label{All anomalies}
 \nonumber
&&\begin{tabular}{|c|c|}
\hline
Anomaly & Phase\\
\hline
\hline
 $\left[U(1)_A\right]^3$ &   $\mbox{dim}_\psi n_\psi q_\psi^3+\mbox{dim}_\chi n_\chi q_\chi^3$ \\
 \hline
$\left[SU(n_\psi)\right]^3$ & $\mbox{dim}_\psi$ \\
 \hline
$\left[SU(n_\chi)\right]^3$  & $\mbox{dim}_\chi$ \\
 \hline
\end{tabular}\,
\begin{tabular}{|c|c|}
\hline
Anomaly & Phase\\
\hline
\hline
 $U(1)_A\left[SU(n_\psi)\right]^2$ &   $q_\psi \mbox{dim}_\psi$ \\
 \hline
 $U(1)_A\left[SU(n_\chi)\right]^2$ &   $q_\chi \mbox{dim}_\chi$ \\
 \hline
 $U(1)_A\left[{\rm grav}\right]^2$ &   $2 \left( q_\psi \mbox{dim}_\psi n_\psi+q_\chi \mbox{dim}_\chi n_\chi \right)$ \\
 \hline
\end{tabular}\,
\\
&& \begin{tabular}{|c|c|}
\hline
Anomaly & Phase\\
\hline
\hline
 $\mathbb Z_r\left[SU(n_\psi)\right]^2$ &   $\mbox{dim}_\psi$ mod $r$ \\
 \hline
 $\mathbb Z_r\left[SU(n_\chi)\right]^2$ &   $\mbox{dim}_\chi$ mod $r$ \\
 \hline
  $\mathbb Z_r\left[U(1)_A\right]^2$ &   $q_\psi^2 \mbox{dim}_\psi n_\psi+q_\chi^2 \mbox{dim}_\chi n_\chi$ mod $r$ \\
  \hline
 $\mathbb Z_r\left[{\rm grav}\right]^2$ &   $2 \left( \mbox{dim}_\psi n_\psi+\mbox{dim}_\chi n_\chi \right)$ mod $r$ \\
 \hline
\end{tabular}\,,
\end{eqnarray}
Notice that if some generators of $\mathbb Z_r$ are redundant (i.e., they can be absorbed in a combination of the color, flavor, and Lorentz centers) such that only $\mathbb Z_p^{d\chi}\subset \mathbb Z_r$ is the genuine discrere chiral symmetry, then we should make the replacement $r\rightarrow p$ in the above tables.

\section{Descending in the CFU Background Fluxes}
\label{Descending in the ACF background fluxes}

We can also calculate the discrete anomaly in the CFU background  using the Stora-Zumino chain of descent procedure. In the following, we assume that we already singled out the good global symmetry that acts on the fermions. In particular, we use the results of Appendix \ref{The descend procedure and counter terms} to read the good part of $U(1)_1\times U(1)_2$ symmetry that survives the quantum corrections: the remaining symmetry is $U(1)_A\times \mathbb Z_r$ (modulo global structure given in eq.~(\ref{eq:global_structure_U(1)_Zr})). Also, we use the results  of Appendix \ref{The descend procedure and counter terms} to read the $U(1)_A$ charges of the fermions.  Moreover, we choose to make use of the canonical counter terms, as in Appendix \ref{The descend procedure and counter terms}. 

  We start as before by writing the anomaly polynomial in $6$ dimensions:
\begin{eqnarray}
\Omega_6=\frac{2\pi}{3!}\sum_{\scriptsize\mbox{Weyl}}\mbox{tr}_{\cal R}\left(\frac{-i{\cal F}_2}{2\pi}\right)^3~,
\end{eqnarray}
where ${\cal F}_2$ is the combined background field strengths of the color and global symmetries and the sum and the trace is taken  over  every  left-handed Weyl fermion transforming in a general representation ${\cal R}$. The explicit form of ${\cal F}_2$ is given in eq. (\ref{full expression}) below.  Integrating the anomaly polynomial over a $6$-dimensional manifold $\mathbb M^6$ gives the phase:
\begin{eqnarray}
e^{i 2\pi \eta}=e^{i\frac{2\pi}{48\pi^3}\sum_{\scriptsize\mbox{Weyl}}\int_{\mathbb M^6}\mbox{tr}_{\cal R}\left[{\cal F}_2\wedge {\cal F}_2\wedge {\cal F}_2 \right]}~,
\label{eta invariant}
\end{eqnarray}
which reproduces the $4$-dimensional anomaly via the descend procedure\footnote{The alert reader will recognize that we chose the Greek letter $\eta$ to denote the phase since this is exactly the $\eta$-invariant of the  Atiyah-Patodi-Singer index theorem \cite{Atiyah:1975jf}.}. In the following we compute the contribution from a single Weyl fermion to $\eta$, while the contribution from other fermion species follows the exact same procedure. In order to reduce notational clutter, we take the flavor group to be $SU(n_f)$, the fermion charge under the axial $U(1)_A$ to be $q$, and the discrete symmetry to be $\mathbb Z_r$.  The calligraphic field strength ${\cal F}_2$ is a superposition of the color $f_2^c$, flavor $F^f_2$, axial $F_2$, and discrete $F^{\mathbb Z_r}_2$ gauge fields:
\begin{eqnarray}
{\cal F}_2=f^{c}_2\otimes \mathbf{1}_f+ \mathbf{1}_c \otimes F^{f}_2 + \mathbf{1}_c\otimes \mathbf{1}_f \otimes \left(qF_2+F^{\mathbb Z_r}_2\right)~,
\label{full expression}
\end{eqnarray}
where $\mathbf{1}_c$ and $\mathbf{1}_f$ are unit matrices of dimensions $\mbox{dim}_{\cal R}$  and $n_f$, respectively. 

In our construction we introduce the gauge field $A^{\mathbb Z_r}_1$, which can be thought of an abelian $U(1)$ gauge field along with a charge-$r$ periodic scalar  $\phi$ satisfying the constraint $d\phi=r A^{\mathbb Z_r}_1$, and hence, $r d A^{\mathbb Z_r}_1= rF^{\mathbb Z_r}_2=0$ mod $r$, while the integral of $d\phi$ over a $1$-cycle is $2\pi$ periodic $\oint d\phi\in 2\pi \mathbb Z$, and hence, $\oint A^{\mathbb Z_r}_1\in \frac{2\pi}{r}\mathbb Z$. The constraint $d\phi=r A^{\mathbb Z_r}_1$ is also invariant under the large gauge transformation $\phi\rightarrow \phi+r\zeta$ and $A^{\mathbb Z_r}_1\rightarrow A^{\mathbb Z_r}_1+d\zeta$ such that $\oint d\zeta\in 2\pi \mathbb Z$. 

 One also needs to turn on  fluxes in the centers of the color, flavor, and $U(1)_A$ directions. Here, we follow exactly the discussion in the bulk of this paper, which we reproduce here in a slightly modified fashion. Let us take the color direction as an example, whereas turning on a center flux in the flavor direction follows the same steps. To this end, we consider a pair of $1$-form and $2$-form fields $\left(B^{c}_2, B^{c}_2 \right)$ such that the constraint $dB^{c}_1=NB^{c}_2$ is obeyed and $\frac{dB^{c}_1}{2\pi}$ over $2$-cycles has integral period: $\oint dB^{c}_1\in 2\pi \mathbb Z$, and therefore, $\oint B^{c}_2\in\frac{2\pi}{N}\mathbb Z$. Next, we enlarge $SU(N)$ to $U(N)$ and embed $B^{c}_1$ into the $U(1)$ factor of $U(N)$. The $U(N)$ field strength is $\hat f^{c}_2=f^{c}_2+\frac{1}{N}dB^{c}_1\mathbf {1}_{c, N}$ (notice that the unit matrix $\mathbf {1}_{c, N}$ has dimension $N$) and it satisfies $\tr_{\Box}\hat f^{c}_2=dB^{c}_1$.  In fact, enlarging the gauge group introduces a spurious extra degree of freedom, which can be eliminated by postulating the following invariance under a  $1$-form gauge field $\lambda^{c}_1$: $\hat f^{c}_2\rightarrow \hat f^c{}+d\lambda^{c}_1$. Subsequently, the pair $\left(B^{c}_1, B^{c}_2 \right)$ transforms as $B^{c}_2\rightarrow B^{c}_2+d\lambda^{c}_1$ and $B^{c}_1\rightarrow B^{c}_1+N\lambda^{c}_1$, which leaves the constraint $dB^{c}_1=NB^{c}_2$ intact. Therefore, we demand that all gauge-invariant quantities are functions of the combination $\hat f^{c}_2-B^{c}_2 \mathbf{1}_{c,N}$, which is left invariant under the postulated  $1$-form gauge field. 

Similarly, we introduce the pair $\left(B^{f}_1, B^{f}_2 \right)$ with $\oint B^{f}_2\in\frac{2\pi}{n_f}\mathbb Z$ in order to turn on a fractional flux in the flavor group.  Then, the total field strength is written as:
\begin{eqnarray}\label{total F}
\begin{split}
{\cal F}=\left[\hat f^{c}_2-\,B^{c}_2 \mathbf{1}_{c, N}\right]\otimes\mathbf{1}_f+ \mathbf{1}_{c}
\otimes\left[\hat F^{f}_2-B^{f}_2 \mathbf{1}_f\right]
+ \mathbf{1}_{c}\otimes \mathbf{1}_f \otimes \left(q (F_2-B_2)+F^{\mathbb Z_r}\right)~,
\end{split}
\end{eqnarray}
keeping in mind that $F^{\mathbb Z_r}=0$ mod $r$ and the $U(1)_A$ background field $B_2$ has to be chosen such that the cocycle conditions are satisfied, see eqs. (\ref{eq:detailed cocycles psi}) and (\ref{eq:detailed cocycles chi}). Next, we substitute (\ref{total F}) into (\ref{eta invariant}) and take the trace in ${\cal R}$. In doing so, we first note that there will be a cubic color term $\sum_{\scriptsize\mbox{Weyl}}\mbox{tr}_{\cal R}\left[\hat f^{c}_2-\frac{1}{N}dB^{c}_1\mathbf{1}_{c, N}\right]^3\sim \sum_{\scriptsize\mbox{Weyl}}\mbox{tr}_{\cal R}\left[T^aT^bT^c\right]$, where ${T^a}$ are the Lie-algebra generators of the color group. This term is zero by construction since our UV theory is gauge-anomaly free. In addition, there is a mixed term between $U(1)_A$ axial and the color fields: $\sum_{\scriptsize\mbox{Weyl}} q (F_2-B_2)\mbox{tr}_{\cal R}\left[\hat f^{c}_2-\frac{1}{N}dB^{c}_1\mathbf{1}_{c,N}\right]^2\sim  \sum_{\scriptsize\mbox{Weyl}}q T_{{\cal R}}$, where $T_{{\cal R}}$ is the Dynkin index of the color group. Again, this term vanishes on group-theoretical grounds since $U(1)_A$  is a good symmetry in the color background.  This leaves us with the following terms:
\begin{eqnarray}
\nonumber
2\pi\eta&=&\frac{1}{24\pi^2}\sum_{\scriptsize\mbox{Weyl}}\int_{\mathbb M_6}\left\{3\,\mbox{dim}_{\cal R}\left(q (F_2-B_2)+F^{\mathbb Z_r}_2\right) \wedge \left[\mbox{tr}_\Box\left(\hat F^{f}_2\wedge \hat F^{f}_2 \right)-n_f B^{f}_2\wedge B^{f}_2\ \right]\right.
\\[3pt]
\nonumber
&&\left.+3n_f T_{\cal R}F^{\mathbb Z_r}_2 \wedge \left[\mbox{tr}_\Box\left(\hat f^{c}_2\wedge \hat f^{c}_2 \right)-NB^{c}_2\wedge B^{c}_2\ \right] + 3n_f\, \mbox{dim}_{\cal R} F^{\mathbb Z_r}_2 \wedge \left(q(F_2-B_2)\right)^2\right.
\\[3pt]
&&\left.+n_f\, \mbox{dim}_{\cal R}\left[q(F_2-B_2)\right]^3+\mbox{dim}_{\cal R}\mbox{tr}_\Box \left[\hat F^{f}_2-fB^{f}_2\right]^3 \right\}~.
\label{full eta invariant}
\end{eqnarray}

Now, we descend from $6$ to $5$ dimensions. For convenience, we define the topological charge densities:
\begin{eqnarray}
\nonumber
&& q_{c} = \frac{1}{8\pi^2}\left[\mbox{tr}_\Box\left(\hat f^{c}_2\wedge \hat f^{c}_2 \right)-NB^{c}_2\wedge B^{c}_2\right]
\,, \nonumber 
\\[3pt]
&& q_{f} = \frac{1}{8\pi^2}\left[\mbox{tr}_\Box\left(\hat F^{f}_2\wedge \hat F^{f}_2 \right)-n_fB^{f}_2\wedge B^{f}_2\right]~,
\\[3pt]
&& q_u = \frac{1}{8\pi^2} \left( F_2 - B_2 \right)^2 \nonumber~.
\end{eqnarray}
Notice, as a side note, that by using  the integrality of the second Chern character on a 4-manifold, e.g., 
\begin{eqnarray}
 \frac{1}{8\pi^2}\int_{\mathbb M^4} \left\{  {\rm tr}_\Box\left( \hat{f}_2^c \wedge \hat{f}_2^c \right) - {\rm tr}_\Box \left( \hat{f}_2^c \right) \wedge {\rm tr}_\Box \left( \hat{f}_2^c \right)\right\}\in \mathbb Z~,
 \end{eqnarray}
we recover the fractional topological charges eqs. (\ref{eq:Q_c}) to (\ref{eq:Q_u}).  

Then, we can read the mixed anomalies directly from (\ref{full eta invariant}). Let us consider, for example, the mixed anomaly of $\mathbb Z_r$ in the CFU background. Collecting the  $F^{\mathbb Z_r}$ terms and descending from $6$ to $5$ dimensions we find\footnote{In fact, upon descending from $6$ to $5$ dimensions we can also add Bardeen counter terms. Potentially, these terms can be used to eliminate non genuine anomalies. As we checked in Appendix \ref{The descend procedure and counter terms},  these counter terms can be set to $0$ as a canonical choice.}:
\begin{eqnarray}
2\pi\eta &&\supset \sum_{\scriptsize\mbox{Weyl}} \int_{\mathbb M^5} A^{\mathbb Z_r}_1 \wedge \left[ \mbox{dim}_{\cal R}\,q_f+n_f T_{\cal R}q_c+ q^2n_f\,\mbox{dim}_{\cal R} q_u \right]~.
\label{anomaly inflow of Zr}
\end{eqnarray}
This is exactly the anomaly inflow action (\ref{eq:inflow_action_Z_p}). Now, as we flow from $5$ to $4$ dimensions we obtain the anomaly
\begin{eqnarray}
 e^{i\frac{2\pi}{r}\sum_{\scriptsize\mbox{Weyl}}\left[ \mbox{dim}_{\cal R}\,Q_f+n_f T_{\cal R}Q_c+ n_f\,\mbox{dim}_{\cal R}q^2 Q_u\right]}~,
\label{eta discrete}
\end{eqnarray}
where
\begin{eqnarray}
Q_c= m_1 m_2 \left( 1-\frac{1}{N} \right)\,, \quad Q_f= p_1 p_2 \left(1 -\frac{1}{n_f} \right)\,,\quad Q_u=(n_1-s_1) (n_2 - s_2)~,
\end{eqnarray}
where $n_{1,2}\in \mathbb Z$, $m_{1,2} \in \mathbb \mathbb{Z}_{N/{\rm gcd}(N,2)}$, $p_{1,2} \in \mathbb Z_{n_f}$, and $s_{1,2}$ denote $U(1)_A$ background fluxes that satisfies the cocycle conditions. 
The expression between brakets is the $4$-dimensional Dirac index in the CFU flux background, and thus, we immediately recover the $\mathbb Z_r \left[ {\rm CFU} \right]$ mixed anomaly. Since $\mathbb Z_{r}$ is a good symmetry of the theory, one might be tempted to conclude that $\sum_{\scriptsize\mbox{Weyl}} n_f T_{\cal R}=0$ on group-theoretical grounds, and thus, slash out the term that multiplies $Q_c$.  However, one simply cannot do that in the case of the discrete symmetry $\mathbb Z_r$ since this anomaly is matched mod $r$.

Similarly, we can collect the $q(F_2-B_2)$ terms to obtain the anomaly inflow action
\begin{eqnarray}\label{eta cont}
\begin{split}
&&2\pi\eta \supset  \sum_{\scriptsize\mbox{Weyl}}\int_{\mathbb M^5} q A_1 \wedge \left[ \mbox{dim}_{\cal R}\,q_f+ q^2 n_f\,\mbox{dim}_{\cal R} q_u\right]
-\frac{q^3n_f\,\mbox{dim}_{\cal R}}{3}\sum_{\scriptsize\mbox{Weyl}}\int_{\mathbb M^5} A_1 \wedge q_u~.
\end{split}
\end{eqnarray}
The first term in the above expression gives the $U(1)_A \left[ {\rm CFU} \right]$ mixed anomaly of (\ref{eq:inflow_action_U(1)}), while the second  term  accounts for the fact that descending from $6$ to $4$ dimensions gives the consistent cubic anomaly, which is accompanied by an extra factor of $\frac{1}{3}$ compared to the covariant anomaly. We must emphasize, however, that anomaly matching works irrespective of whether we are computing the consistent or covariant anomalies. 

One important difference between (\ref{anomaly inflow of Zr}) and (\ref{eta cont}) is that the color sector contributes a fractional flux to the discrete anomaly, while the continuous anomaly, as we pointed above, is totally blind to the color sector. This observation plays a pivotal role in matching the anomalies in the IR, as we discussed in the bulk.

\section{The 3-loop $\beta$-Function and IR Fixed Points}
\label{2loop beta function}

The 3-loop $\beta$ function is given by (see \cite{Caswell:1974gg,Dietrich:2006cm,Zoller:2016sgq})
\begin{eqnarray}
\begin{split}
\beta(g)=&-\beta_0\frac{g^3}{(4\pi)^2}-\beta_1\frac{g^5}{(4\pi)^4}-\beta_2\frac{g^7}{(4\pi)^6}~,
\\[3pt]
\beta_0=&\frac{11}{6}C_2(G)-\sum_{\cal R}\frac{1}{3}T_{\cal R}n_{\cal R}~,
\\[3pt]
\beta_1=&\frac{34}{12}C_2^2(G)-\sum_{{\cal R}}\left\{\frac{5}{6}n_{\cal R} C_2(G)T_{\cal R} + \frac{n_{\cal R}}{2}C_2({\cal R})T_{\cal R}\right\}~,
\\[3pt]
\beta_2=&\frac{2857}{432}C_2^3(G)-\sum_{\cal{R}}\frac{n_{\cal R}T_{\cal R}}{4} \left[-\frac{C_2^2({\cal R})}{2}+\frac{205 C_2(G)C_2({\cal R})}{36}+\frac{1415C_2^2(G)}{108} \right]
\\[3pt]
&+\quad\sum_{\cal{R},\cal{R}'} \frac{n_{\cal R}n_{\cal R}' T_{\cal R}T_{\cal R'}}{16}\left[\frac{44 C_2({\cal R})}{18}+\frac{158C_2(G)}{54} \right]~,
\end{split}
\label{beta function}
\end{eqnarray} 
where $G$ denotes the adjoint representation and $n_{\cal R}$ is the number of the Weyl flavors in representation ${\cal R}$. The quadratic Casimir operator of representation ${\cal R}$,  $C_2({\cal R})$, is defined as 
\begin{eqnarray}
t^a_{\cal R}t^a_{\cal R}=C_2({\cal R})\mathbf{1}_{\cal R}~,
\label{Casimir}
\end{eqnarray}
and $C_2(G)$ is the quadratic Casimir of the adjoint representation. $T_{\cal R}$ is the Dynkin index in the same representation which is defined by
\begin{eqnarray}
\mbox{tr}\left[t^a_{\cal R}t^b_{\cal R}\right]=T_{\cal R}\delta^{ab}~.
\label{trace}
\end{eqnarray}
From Eqs. (\ref{Casimir}) and (\ref{trace}) we easily obtain the useful relation 
\begin{eqnarray}
T_{\cal R}\mbox{dim}_G=C_2({\cal R})\mbox{dim}_{\cal R}~,
\end{eqnarray}
where $\mbox{dim}_{\cal R}$ is the dimension of $\cal R$.

In particular, we have $C_2(G)=2N$, $\mbox{dim}_G=N^2-1$, $T_\psi=N+2$, $\mbox{dim}_\psi=\frac{N(N+1)}{2}$, $C_2(\psi)=\frac{2(N+2)(N-1)}{N}$, $T_\chi=N-2$,  $\mbox{dim}_\chi=\frac{N(N-1)}{2}$, $C_2(\chi)=\frac{2(N-2)(N+1)}{N}$. Then, the values of $\beta_0$ to $\beta_2$ are 
\begin{eqnarray}
\begin{split}
\beta_0=&\ \frac{1}{3}\left[11N-\frac{2}{k}(N^2-8) \right]~,
\\[3pt]
\beta_1=&\ \frac{2\left(-48+76N^2+17kN^3-8N^4\right)}{3kN}~,
\\[3pt]
\beta_2=&\ \frac{1}{54k^2N^2}\Big [ 2857k^2N^5+N(-8448+12448N^2-2584N^4+145N^6)
\\[3pt]
&\quad -2k(864+3948N^2-8945N^4+988N^6)\Big ]~.
\end{split}
\end{eqnarray}
Assuming that $\beta_0>0$ and $\beta_1<0$, then the theory develops an IR fixed point to $2$-loops. The value of the coupling constant at the fixed point is
\begin{eqnarray}
\alpha_{\scriptscriptstyle \rm IR}\equiv \frac{g_{\scriptscriptstyle \rm IR}^2}{4\pi}=-\frac{4\pi \beta_0}{\beta_1}=\frac{2\pi N \left(16+11kN-2N^2\right)}{48-76N^2-17kN^3+8N^4}~.
\end{eqnarray}
The robustness of this fixed point can be checked by finding the roots of the $\beta$-function after including the $3$-loop term.

\bibliography{References.bib}
  
\bibliographystyle{JHEP}

\end{document}